\documentclass[prb,twocolumn,amsmath,amssymb,superscriptaddress,nobalancelastpage]{revtex4}
\usepackage{graphicx}
\usepackage{ulem}
\usepackage{color}
\usepackage{graphicx}  
\usepackage{dcolumn}   
\usepackage{bm}        
\usepackage{amssymb} 
\usepackage{pdfpages}

\newcommand{\be}{\begin{eqnarray}}
\newcommand{\ee}{\end{eqnarray}}

\usepackage{mathtools}

\DeclarePairedDelimiter\ket{\lvert}{\rangle}
\DeclarePairedDelimiterX\braket[2]{\langle}{\rangle}{#1 \delimsize\vert #2}

\newcommand{\bfk}{\mathbf{k}}

\newcommand{\bfq}{\mathbf{q}}
\newcommand{\bfQ}{\mathbf{Q}}

\usepackage{enumerate}
\usepackage{epstopdf}
\usepackage{amsmath}

\newcommand{\Tr}{\mathrm{Tr}}

\newcommand{\sgn}{\mathrm{sgn}}

\begin{document}

\title{Scaling of the chiral magnetic effect in quantum diffusive Weyl semimetals}
\author{Yen-Ting Lin}
\affiliation{Department of Physics, National Tsing Hua University, Hsinchu 30043,
Taiwan, 300, R.O.C.}
\author {Liang-Jun Zhai}
\affiliation{The School of Mathematics and Physics, Jiangsu University of Technology, Changzhou 213001, China}
\author {Chung-Yu Mou}
\affiliation{Department of Physics, National Tsing Hua University, Hsinchu 30043,
Taiwan, 300, R.O.C.}
\affiliation{Institute of Physics, Academia Sinica, Nankang, Taiwan, R.O.C.}
\affiliation{Physics Division, National Center for Theoretical Sciences, Hsinchu, Taiwan, R.O.C.}

\date{\today}

\begin{abstract}
We investigate the effect of short-range spin-independent disorder on the chiral magnetic effect (CME) in Weyl semimetals. 
Based on a minimum two-band model, the disorder effect is examined 
in the quantum diffusion limit by including the Drude correction and the correction due to
the Cooperon channel. It is shown that the Drude correction
renormalizes the CME coefficient by a factor to a finite value that is independent of
the system size. Furthemore, due to an additional momentum expansion involved in
deriving the CME coefficient, the contribution of Cooperon to  the CME coefficient is governed 
by the quartic momentum term. As a result, in contrast to the weak localization and weak anti-localization effects observed in 
the measurement of conductivity of Dirac fermions, we find that in the limit of zero magnetic field,
the CME coefficients of finite systems manifest the same scaling of localization even in three dimension. 
Our results indicate that while the chiral magnetic current due to slowly oscillating magnetic fields can exist in clean systems, 
its observability will be limited by suppression due to short-range disorder in condensed matters.
\end{abstract}
\maketitle

\section{Introduction}

The discoveries of graphene\cite{Geim1,Geim2} and topological insulators\cite{Kane, Zhang} have 
revived the interest of simulating relativistic particles in condensed matter systems.  After intensive studies in
the past few years, it is now realized that in 3D, materials in proximity to topological insulating phases may exhibit relativistic semi-metallic phases\cite{Murakami,Chou}. Particularly, in the presence of the time-reversal and inversion symmetries,
3D Dirac semi-metals with  four-fold degeneracy are realized as long as certain crystalline symmetries are supplemented\cite{Na3Bi, Cd2As3_1,Cd2As3_2}. The underlying quasi-particles in these materials simulate Dirac fermions without definite handness (or chirality). It is further demonstrated recently in TaAs\cite{TaAs1,TaAs2} that by breaking either the inversion symmetry or the time-reversal symmetry, the long-sought analogy of Weyl fermions with definite chirality can be realized in condensed matters. 

The Weyl semimetal is a new phase of materials in which the energy dispersion of quasi-particles support nodal
points that result from splitting of Dirac nodes either through the breaking of time-reversal symmetry or inversion symmetry\cite{review, Zhai}.
Due to the Nielsen-Nimomiya theorem\cite{Nielsen}, chiral fermions on lattice can not exist alone. The net chirality due to all Weyl nodes  must vanish. Hence Weyl nodes must occur in pairs with opposite chiralities. The non-vanishing chirality for each Weyl node is the main source of the novel properties proposed in the past 
for Weyl semimetals, such as anomalous Hall effect and chiral magnetic effect (CME)\cite{Burkov2015}. In particular, it is known that while  the total chirality due to all Weyl nodes  must vanish, the chiral current in a Weyl semimetal is not conserved. This is well-known as the chiral anomaly\cite{Burkov2012}. One of the peculiar consequences due to the chiral anomaly is the prediction of a non-dissipative persistent current, $\mathbf{J}$, in parallel to the magnetic field when the Weyl semimetal is placed in a static magnetic field, $\mathbf{B}$, i.e., $\mathbf{J}=\alpha\mathbf{B}$. Here $\alpha$ is  known as the CME coefficient or chiral magnetic conductivity and is proportional to the energy separation between Weyl nodes.
While in the continuum model of Weyl Fermions, the CME effect is generally found when an ultraviolet energy cutoff 
for the linear energy dispersion is introduced. In condensed matters on lattice, the cutoff is in the momentum space\cite{two-band}. As a result, the CME effect is generally absent in equilibrium condensed matter systems\cite{two-band,Basar, Franz}.
Indeed, according to a general argument\cite{Franz}, the existence of a nonzero equilibrium chiral magnetic current induced by a static magnetic field at zero temperature would imply that one can extract energy from ground state\cite{Franz}. A no-go theorem is also established showing that the equilibrium ground state of a given system in the thermodynamic limit would not be able to carry any current\cite{Yamamoto}. Hence in static magnetic fields, the ground state of condensed matter systems would not support the CME effect. 

On the other hand, according a derivation based on the Kubo formula in the uniform limit, i.e., $\mathbf{q}=0$ before
the frequency  $\omega \rightarrow 0$\cite{two-band}, one obtains a non-vanishing electric current in parallel the oscillating magnetic field.   Hence the CME effect exists in a slowly oscillating magnetic field that prevents equilibration of the system\cite{Scattering theory,gyro, gyro1, gyro2, gyro3}. More precisely, the CME current is in proportional to the AC magnetic field, $\mathbf{B} (t)$ with $\mathbf{J} (t) =\alpha\mathbf{B}(t)$. Here the CME coefficient $\alpha$ may depend the DC magnetic field if in addition to
the AC field, a static magnetic field is applied. Note that for non-topological contribution, the magnetic induced current has been termed as the gyrotropic magnetic effect\cite{gyro} and the CME coefficient is related to intrinsic magnetic moment. For Weyl semimetals on lattice, the chiral magnetic current can be accounted by non-vanishing Berry curvature when the energy dispersion deviates from the linear dispersion\cite{two-band}. 

While the CME effect may exist for condensed matters in slowly oscillating magnetic fields, so far, 
all of the considerations focus on clean systems. Real systems inevitably include disorder or impurities. 
It is thus necessary to examine the effect of disorder on the CME effect. Conventionally, because the chiral 
magnetic current is due to the imbalance of chirality and the chirality of a Weyl point is topologically stable against 
perturbations,  it is usually stated without proof that the chiral magnetic current is topologically protected. However, 
from the study of effects of disorder on conductivity, it is known that in addition to the suppression of electric current due to momentum scatterings by short-range disorder, interference due to quantum diffusion is also in presence and can lead to  the phenomena of weak localization (WL) or weak anti-localization (WAL) that coexist with the phenomenon of negative resistance\cite{neg, Hikamibox, ShenPRB2015, MacDonald2014}. Hence it is crucial to examine the interference effect on the CME effect due to quantum diffusion induced by short-range disorder.

In this work, based on a minimum two-band model of the Weyl semimetal, we examine the disorder effect 
in the quantum diffusion limit by extending the analysis on conductivity due to short-range disorder.
In particular, we include the Drude correction and correction to the CME effect due to
the Cooperon channel.  We find that while the Drude correction
renormalizes the CME coefficient to a finite value, the dominant 
scaling contribution (in terms of system size $L$)  to the correction of the CME coefficient 
is due to the Cooperon channel. Furthermore, we find that instead of making contribution through diffusive propagation, 
the contribution of Cooperon to the 
CME coefficient is governed by the quartic momentum term.  As a result, the finite size scaling of CME coefficient differs from that of the conventional conductivity. We find that for intra-Wey-node channel, localization dominates so that the CME coefficients of finite systems manifest the same scalings as  that in three dimension. Our results indicate that disorder tends to suppress the chiral magnetic current that results from applied slowly oscillating magnetic fields on Weyl semimetals. The observability of the CME effect is thus limited by suppression due to disorders in condensed matters.

This paper is organized as follows: In Sec.\uppercase\expandafter{\romannumeral2}, we describe the minimal two band model with short-range disorder for Weyl semimetals. The computation of the CME coefficient due to short-range disorder is formulated. In Sec.\uppercase\expandafter{\romannumeral3}, the contribution to the CME coefficient in quantum diffusion limit is formulated. In Sec.\uppercase\expandafter{\romannumeral4},  detailed calculations of the quantum diffusive correction to the CME coefficient are presented. Finally, in Sec.\uppercase\expandafter{\romannumeral5}, we conclude and discuss the relevancy of our results to experiments.

\section{Theoretical Formulation of Chiral Magnetic Effect} \label{sec2}
We start by considering a generic minimum model for the Weyl semimetals with the Hamiltonian being given by 
\begin{eqnarray} \label{Hami}
H = \sum_{\mathbf{k} \sigma \sigma'}  h^{\sigma \sigma'} ({\mathbf k}) c_{\mathbf{k} \sigma}^\dag c_{\mathbf{k} \sigma'}+ H.C.. \label{hh}
\end{eqnarray}
Here $h({\mathbf k})$ is a $2\times 2$ matrix which can be generically expressed in the form $h (\mathbf{k})=\epsilon(\mathbf{k})+ \mathbf{d}(\mathbf{k})\cdot\bm{\sigma}$ characterized by $\epsilon(\mathbf{k})$ and $\mathbf{d}(\mathbf{k})$, where $\sigma^{i}$ ($\i =x,y,z$) are three Pauli  matrices and $\mathbf{k}$ is the Bloch wave-vector for the electron creation and annihilation operators $c_{\mathbf{k} \sigma}^\dag$ and $c_{\mathbf{k} \sigma}$. The band energies for the generic model Eq.(\ref{hh}) are given by
\begin{equation}
E_s (k) = \epsilon(\mathbf{k})+s d(\mathbf{k}),
\end{equation}
where $s=\pm$ and $d(\mathbf{k})$ is the magnitude of the vector $\mathbf{d}(\mathbf{k})$.
In the simplest realization of the Weyl semimetal, one extends the Qi-Wu-Zhang (QWZ) two-band model 
on a cubic lattice to include two Weyl nodes. The Hamiltonian $h({\mathbf k})$ can be written as \cite{two-band, tewari}
\begin{equation}
h (\mathbf{k})=t_z\cos k_z+h_{so}+h_m
\end{equation}
with $h_{so}$ and $h_m$ being given by 
\begin{eqnarray}
&&h_{so}(\mathbf{k})=\lambda_{so}\left( \sin k_x \sigma^x+\sin k_y \sigma^y+\sin k_z \sigma^z \right) \\
&&h_m(\mathbf{k})=\left( m+2-\cos k_x-\cos k_y \right) \sigma^z \; .
\label{eq:model}  
\end{eqnarray}
It is important to note that the hopping along the z-direction, $t_z\cos k_z$, is the key term that 
controls the chiral magnetic effect. In this model, the inversion symmetry is broken by the spin-orbit interaction $h_{so}$, while the time-reversal symmetry is broken by $h_m$. Breaking one of these symmetries leads the separation of 
the Dirac node into two Weyl nodes.  We shall focus on the case of $m=0$, in which the Weyl nodes emerge at $\mathcal{C}^{+}=(0,0,0)$ and $\mathcal{C}^{-}=(0,0,\pi)$\cite{two-band}. Since the hopping term  $t_z$ leads to an energy change of $t_z \cos k_z$, it is clear that $t_z$ controls the relative shifts of Weyl nodes at $(0,0,0)$ and $(0,0,\pi)$ in energy and thus controls the chirality imbalance that leads to the chiral magnetic effect. 
Specifically, the linearized Hamiltonian around nodal points, $\mathcal{C}^{+}=(0,0,0)$ and $\mathcal{C}^{-}=(0,0,\pi)$, is given by 
\begin{eqnarray}
{h}(\mathbf{k})=\pm t_{z}+v_F \mathbf{k}_{\pm}\cdot\bm{\sigma},\label{eq:linearized QWZ model}
\end{eqnarray}
where $\mathbf{k}_{+}=(k_x,k_y,k_z)$ and $\mathbf{k}_{-}=(k_x,k_y,-k_z)$ and $v_F=\lambda_{so}$ is the Fermi velocity. Hence when the Fermi energy $\epsilon_F$ is zero, the energies of two Weyl points relative to $\epsilon_F$ are $\pm t_z$.

In the presence of electromagnetic fields, the linear response of the induced current can be generally expressed as
\begin{equation}
J^i(\mathbf{q},\omega) = \Pi_{ij}(\mathbf{q},\omega)
A^j(\mathbf{q},\omega), \label{JJ}
\end{equation}
where $\Pi_{ij}$ is the  retarded current-current correlation function. On the other hand, the
chiral magnetic effect is the induced electric current in parallel to the applied magnetic field and can be expressed as 
\begin{eqnarray}
J^i_{{\rm CME}}(\mathbf{q},\omega)
&=& \alpha(\mathbf{q},\omega) B^i(\mathbf{q},\omega) \nonumber \\
&=& -i \alpha(\mathbf{q},\omega) \epsilon^{ijk} q_k A^j(\mathbf{q},\omega),  \label{CMEJJ}
\end{eqnarray}
where $\alpha$ is the CME coefficient. By comparing Eqs.(\ref{JJ}) and (\ref{CMEJJ}), it is clear that 
because the Levi-Civita tensor $\epsilon^{ijk}$ is antisymmetric in $i$ and $j$, the CME coefficient is determined by
the anti-symmetric part of $\Pi_{ij}$\cite{two-band,tewari}. Hence by choosing $z$ as the direction of the current, the CME coefficient is determined by the anti-symmetric part of $\Pi_{xy}$\cite{two-band,tewari}. After averaging over three directions of current, the CME coefficient can be expressed as 
\begin{eqnarray}
\alpha(\mathbf{q},\omega)=\frac{i}{3q} (\Pi_{xy}^{{\rm anti}}(q\hat{z},\omega) + \Pi_{yz}^{{\rm anti}}(q\hat{x},\omega) + \Pi_{zx}^{{\rm ant}i}(q\hat{y},\omega)). \nonumber \\
\end{eqnarray}
In the clean limit and in a slowly oscillating magnetic field, the CME coefficient is given by\cite{two-band}
\begin{eqnarray}\label{chmag1}
&&\lim_{\omega\to 0} \lim_{q\to 0} \alpha(\mathbf{q},\omega) \nonumber\\
&&=\frac{e^2}{\hbar}\frac{1}{\Omega}\sum_{\bfk} \; \sum_{t=\pm} \left[\frac{\mathbf{v}_{\mathbf{k},+}+\mathbf{v}_{\mathbf{k},-}}{2} \cdot \mathbf{\Omega}_{\mathbf{k},t} \; n_{F}^{0(t)}(\mathbf{k}) \right. \nonumber\\%
&&\qquad\qquad\qquad \left. %
-\frac{1}{3} t\;d(\mathbf{k}) \; \mathbf{v}_{\mathbf{k},t} \cdot \mathbf{\Omega}_{\mathbf{k},t} \; \frac{\partial n_{F}^{0(t)}(\mathbf{k})}{\partial E_t}\right] \; ,
\end{eqnarray}
where $\mathbf{\Omega}_{\pm}$ are Berry curvatures and $n_{F}^{0(t)}$ is the Fermi-Dirac distribution function for energy bands $E_t$. In particular, one finds that $\alpha>0$ for $t_z >0$ and $\epsilon_F=0$ \cite{two-band}. 

To include effects due to disorder, it is necessary to perform the average over disorder. We shall assume that the randomly distributed impurities are characterized by a Gaussian ensemble with a potential described by 
$\langle V(\mathbf{r}) \rangle=0$ and short-range correlation
\begin{eqnarray}
\langle V(\mathbf{r})_{\alpha \gamma} V(\mathbf{r'})_{\beta \delta} \rangle=\gamma_{e}b^{\alpha\gamma}_{\beta \delta} \delta (\mathbf{r}-\mathbf{r'} ).
\label{impurity correlation}
\end{eqnarray}
Here $\gamma_e$ characterizes the strength of disorder. For spin-independent potential impurities, $b^{\alpha\gamma}_{\beta \delta}$ is a unit $4 \times 4$ matrix. The perturbation theory can be developed to compute the disorder-averaged physical quantities. After averaging over disorder, the translational invariance is restored and the CME coefficient is thus given by
\begin{eqnarray}
\alpha(\mathbf{q},\omega)=\frac{i}{3q}  \langle \Pi_{xy}^{{\rm anti}}(q\hat{z},\omega) + \Pi_{yz}^{{\rm anti}}(q\hat{x},\omega) + \Pi_{zx}^{{\rm anti}}(q\hat{y},\omega) \rangle . \nonumber \\
\end{eqnarray}
\section{\protect\bigskip Quantum Diffusive Correction}
To include the effect of disorder, we first note that one of the  important effects due to disorder observed in 
the experiments on conductivity is the suppression or enhancement of the electric current via quantum 
interference known as  the phenomena of weak localization or weak anti-localization 
in magnetoresistance\cite{neg, Hikamibox,ShenPRB2015, MacDonald2014}.
Since the Cooperon channel and related leading order channels  are the dominant contribution in the scaling 
behavior of conductivity in the phenomena of weak localization or weak anti-localization,  we shall 
therefore focus on the corrections due to the Cooperon channel in the quantum diffusive regime.  
Other contributions (such as vertex renormalization and Drude  contribution) are also included
along with the derivation of the contribution due to Cooperon .

\begin{figure}[t]
\includegraphics[height=3.4in,width=3.2in] {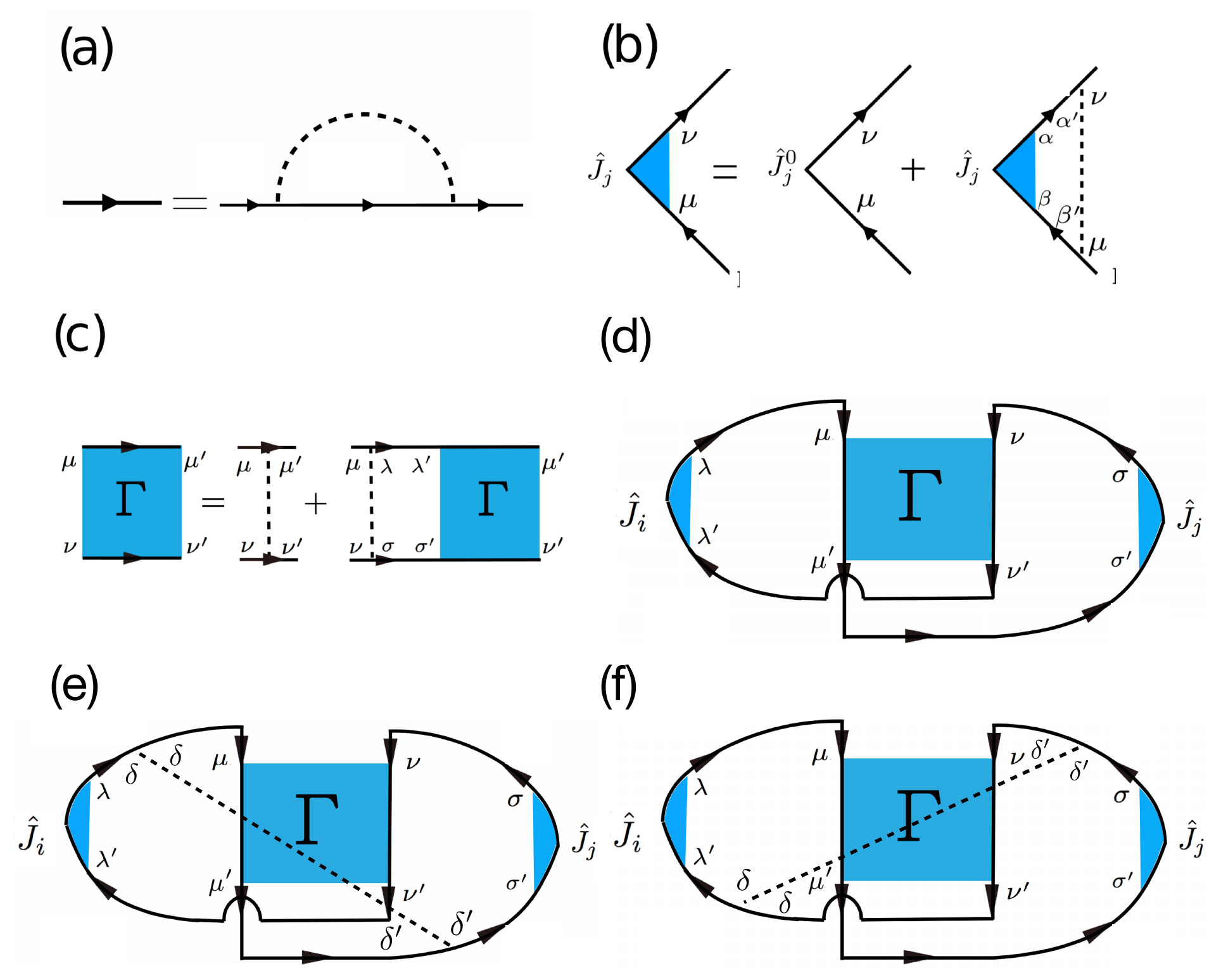} 
\caption{Feymann diagrams for the disorder correction of the CME coefficient. (a) Each solid line represents the Green's function with first order Born approximation. (b) The vertex correction of the CME coefficient. (c) The correction due to the ladder diagram for particle-particle channel with the maximally crossed diagrams. The leading corrections of the current-current correlation due to (d) the bare and (e) \& (f) two dressed Hikami boxes from the maximally crossed diagrams. It turns out that two dressed Hikami boxes cancel each other out due to the anti-symmetric combination of current-current correlation in the CME coefficient (see text for details).}
\label{fig1}
\end{figure}
To find the correction due to the Cooperon channel, we shall start by considering the
correction of the Green's function due to disorder. After averaging over disorder,
the unperturbed retarded Green's function is corrected by self-energy. The corrected Green's function
is represented diagrammatically by a solid line shown in Fig.~\ref{fig1}(a) and is given by
\begin{eqnarray}
&&\hspace{-2.5em}G_R (\mathbf{k},\omega) = \frac{1}{\omega +\epsilon_F-\hat{h}(\mathbf{k})+{i\eta}}\label{Green}, 
\end{eqnarray}
where $\epsilon_F$ is the Fermi energy and  $\eta$ is due to the self-energy correction.  
We shall assume that $\eta$ takes the form given by the first-order Born approximation and is given by\cite{Akkermans}
\begin{eqnarray}
\eta  \equiv \frac{1}{2\tau_e}=-\mathrm{Im}\Sigma^R(\mathbf{k})=\pi N(\bar{\epsilon}_F) \gamma_e ,
\end{eqnarray}
where $\tau_e$ is the life-time of the quasi-particle, $\Sigma^R(\mathbf{k})$ is the self-energy, and 
$N(\bar{\epsilon}_F) $ is the density of states at the Fermi level relative to the Weyl node. When $\epsilon_F=0$, $\bar{\epsilon}_F= \pm t_z$.  One finds that 
\begin{equation}
\eta = \frac{\gamma_e t^2_z}{2 \pi \lambda_{so}^3}.  \label{eta}
\end{equation}

To calculate  $\langle \Pi_{xy}^{{\rm anti}}(\mathbf{q},\omega) \rangle$ perturbatively, three corrections beyond the clean
limits are included in the current-current correlation as shown in Figs.~\ref{fig1}(d), (e) and (f). These corrections are the leading corrections due to the Cooperon channel in computing dc conductivity based on the Kubo-Streda formula\cite{Hikamibox}.
As we shall see, however, for the CME correction, two dressed Hikami boxes (e) and (f) cancel each other out due to the anti-symmetric combination of current-current correlation in the CME coefficient 
In these diagrams, the current operator  $ \hat{J}_i$ 
is corrected with the vertex correction shown in Fig.~\ref{fig1}(b)  and is given by 
\begin{eqnarray}
&&\hspace{-1.1em}[\hat{J}_{j}(\bfk\hspace{-0.3em}+\hspace{-0.3em}\bfq)]_{\nu\mu}=[\hat{J}^{0}_{j}(\bfk)]_{\nu\mu}\nonumber\\
&&\hspace{-1.1em}+\frac{1}{V}\sum_{\bfk'}\left[\mathcal{G}_{\alpha'\alpha}(\bfk')[\hat{J}_{j}(\bfk\hspace{-0.2em}+\hspace{-0.2em}\bfq)]_{\alpha\beta}\mathcal{G}_{\beta \beta'}(\bfk'\hspace{-0.3em}+\hspace{-0.2em}\bfq)\gamma_{e}b^{\alpha'\mu}_{\beta'\nu}\right],\nonumber \\ \label{vertex}
\end{eqnarray}
where $V$ is the volume of the system and $\hat{J}^{0}_{j}(\bfk)$ is the unperturbed current operator given by $\hat{J}^{0}_{j}(\bfk) = - \frac{e}{\hbar} \frac{\partial h(\mathbf{k})}{ \partial \mathbf{k}}$
We shall show that the vertex correction contributes a renormalization of $\hat{J}^{0}$ with $\hat{J}_{i} \propto \hat{J}^{0}_i$ in the low energy limit. 

The central block $\Gamma$ shown in Figs.~\ref{fig1}(d), (e), and (f) is the summation of the particle-particle ladder diagram shown in Fig.~\ref{fig1}(c). This is the Cooperon channel with the amplitude represented by $\Gamma(\bfQ+\bfq)$. Here $\bfQ = \bfk + \bfk'$ is the total momentum of the Cooperon and $\bfq$ is the external momentum coming from the external AC magnetic field.  The first correction (Fig.~\ref{fig1}(d)) due to the Cooperon channel can be expressed in terms of the Cooperon propagator, $\Gamma$, weighted by a weighting factor, $W$, as follows\cite{MacDonald2014}
\begin{widetext}
\begin{eqnarray}
&& \langle \Pi_{xy}^{{\rm anti}}(q \hat{z},i\nu_m) \rangle \nonumber\\
&& =\frac{1}{2 \beta V^2}\sum_{n,\bfk,\bfk'} \left[(\mathcal{G} \hat{J}_x \mathcal{G})_{\nu' \mu}(\bfk ,q \hat{z};i\omega_n,i\nu_m) \Gamma^{\mu\mu'}_{\nu\nu'}(\bfk,\bfk',q \hat{z};i\omega_n,i\nu_m)(\mathcal{G} \hat{J}_y \mathcal{G})_{\mu'\nu}(\bfk' ,q \hat{z};i\omega_n,i\nu_m)  \right]- (x\leftrightarrow y)  \nonumber\\
&& \approx\frac{1}{2 \beta V^2}\sum_{n,\bfk,\bfQ}\left[(\mathcal{G} \hat{J}_x \mathcal{G})_{\nu' \mu}(\bfk ,q \hat{z};i\omega_n,i\nu_m)\Gamma^{\mu\mu'}_{\nu\nu'}(\bfQ,q \hat{z};i\omega_n,i\nu_m) (\mathcal{G} \hat{J}_y \mathcal{G})_{\mu'\nu}(-\bfk ,q \hat{z};i\omega_n,i\nu_m)  \right]- (x\leftrightarrow y)\nonumber\\
&&= \sum_m \Tr \left[ W_{xy} (q \hat{z}, i\nu_m) \frac{1}{V}\sum_{\bfQ} \Gamma_{xy}(\bfQ,q  \hat{z}) \right]. \label{cooperon}
\end{eqnarray}
\end{widetext}
Here $\nu_m$ is the Matsubara frequency.  $\mu,\mu',\nu,\nu'$ are spin indices.  In going from the first equation to the second equation, we
have rewritten $\mathbf{k}$ and $\mathbf{k}'$ as $\mathbf{k}+\mathbf{Q}/2$ and $-\mathbf{k}+\mathbf{Q}/2$ and assumed 
that the contributions of $\mathbf{Q} \sim 0$ and $i \omega_n=0$ dominate so that $\Gamma$ depends only on $\mathbf{Q}$ and $\mathbf{q}$.
$W$ is a weighting factor composed
by four Green functions and current operators 
\begin{eqnarray}
&&W^{\mu'\mu}_{\nu'\nu}(\mathbf{q},i\nu_m)\nonumber = \nonumber \\
&& \frac{1}{2\beta V}\sum_{n,\bfk} \left[(\mathcal{G} \hat{J}_x \mathcal{G})_{\nu' \mu}(\mathcal{G} \hat{J}_y \mathcal{G})_{\mu'\nu}(\bfk,\mathbf{q};i\omega_n,i\nu_m)\right. \nonumber\\
&&\left.-(\mathcal{G} \hat{J}_y \mathcal{G})_{\nu' \mu}(\mathcal{G} \hat{J}_x \mathcal{G})_{\mu'\nu}(\bfk ,\mathbf{q};i\omega_n,i\nu_m)\right],  \label{W}
\end{eqnarray}
while $\Gamma$ is the Cooperon propagator determined by the equation
\begin{eqnarray}
&&\Gamma^{\mu\mu'}_{\nu\nu'}(\bfQ ,\mathbf{q};i\omega_n,i\nu_m)  \nonumber \\
&& =\gamma_e{b}^{\mu\mu'}_{\nu\nu'}\hspace{-0.2em}+\hspace{-0.2em}\gamma_e{b}^{\mu\lambda}_{\nu\sigma}\frac{1}{V}\sum_{\bfk} \left[ \mathcal{G}^{\lambda\lambda'}(\bfk\hspace{-0.2em}+\hspace{-0.2em}\bfQ\hspace{-0.2em}+\hspace{-0.2em}\mathbf{q},i\omega_n\hspace{-0.2em}+\hspace{-0.2em}i\nu_m)\right. \nonumber \\
&&\left. \hspace{5em}\times\mathcal{G}_{\sigma\sigma'}(-\bfk,i\omega_n)\Gamma^{\lambda'\mu'}_{\sigma'\nu'}(\bfQ ,\mathbf{q};i\omega_n,i\nu_m)\right]\label{ladder diagram}.\nonumber \\ \label{cooperon1}
\end{eqnarray}
Since it is expected that $\bf Q \approx 0$ gives the largest contribution, we shall use the linearized two-band model and neglect the frequency dependent in Green functions to capture quantum diffusive behavior of $\Gamma$ in the low energy limit .
Note that both $W$ and $\Gamma$ in Eq. (\ref{cooperon}) are arranged into $4 \times 4$ matrices by using the tensor product basis $\{\ket{\uparrow\uparrow},\ket{\uparrow\downarrow},\ket{\downarrow\uparrow},\ket{\downarrow\downarrow} \}$. 

On the other hand, the contribution due to the dressed Hikami box, Fig.~\ref{fig1}(e), can be expressed as
\begin{widetext}
\begin{eqnarray}
&& \langle \Pi_{xy}^{(e)}(q \hat{z},i\nu_m) \rangle \nonumber\\
&& =\frac{\gamma_e}{2 \beta V^3}\sum_{n,\bfk,\bfk',\bfQ} \left\{ \left[(\mathcal{G}\hat{J}_x \mathcal{G})_{\nu' \delta}(\bfk ,q \hat{z};i\omega_n,i\nu_m)\mathcal{G}_{\delta \mu}(\bfk+q\hat{z}, i\omega_n+i\nu_m) \right. \right. \nonumber \\
&& \hspace{1.5cm} \left. \left. \times \Gamma^{\mu\mu'}_{\nu\nu'}(\bfQ,q \hat{z};i\omega_n,i\nu_m)\mathcal{G}_{\mu' \delta'}(-\bfk+q\hat{z}, i\omega_n+i\nu_m) (\mathcal{G} \hat{J}_y \mathcal{G})_{\delta'\nu}(\bfk' ,q \hat{z};i\omega_n,i\nu_m)  \right]- (x\leftrightarrow y) \right\} \nonumber\\
&&= \sum_m \Tr \left[ \bar{W}^{(e)}_{xy} (q \hat{z}, i\nu_m) \frac{1}{V}\sum_{\bfQ} \Gamma_{xy}(\bfQ,q  \hat{z}) \right]. \label{hikamibox1}
\end{eqnarray}
\end{widetext}
Here $\mu,\mu',\nu,\nu', \delta$, and $\delta'$ are spin indices and summations over repeated indices are taken.  
$\bar{W}^{(e)}$ is the corresponding weighting factor composed
by six Green functions and current operators 
\begin{widetext}
\begin{eqnarray}
&&\bar{W}^{\mu'\mu}_{\nu'\nu}(\mathbf{q},i\nu_m)\nonumber = \nonumber \\
&& \frac{\gamma_e}{2\beta V^2}\sum_{\delta, \delta', n,\bfk,\bfk'} \left[(\mathcal{G} \hat{J}_x \mathcal{G})_{\nu' \delta}(\bfk',q\hat{z},
i\omega_n,i\nu_m)
\mathcal{G}_{\delta \mu} (\bfk+\bfq, i \omega_n+i \nu_m) \mathcal{G}_{\mu' \delta'} (-\bfk'+\bfq, i \omega_n+i \nu_m)
(\mathcal{G} \hat{J}_y \mathcal{G})_{\delta'\nu}(-\bfk,\bfq;i\omega_n,i\nu_m)\right. \nonumber\\
&&\hspace{1cm} \left.-  (x\leftrightarrow y) \right].  \label{Wbar}
\end{eqnarray}
\end{widetext}
Similarly, because the dressed Hikami box, Fig.~\ref{fig1}(f) is the same as the dressed Hikami box, Fig.~\ref{fig1}(e), after  
$\hat{J}_x$ and $ \hat{J}_y$ are exchanged and the wavevector $\bfq$ is reversed, the dressed Hikami box, Fig.~\ref{fig1}(f), contributes
\begin{eqnarray}
&& \langle \Pi_{xy}^{(f)}(q \hat{z},i\nu_m) \rangle \nonumber\\
&& = \sum_m \Tr \left[ \bar{W}^{(f)}_{xy} (q \hat{z}, i\nu_m) \frac{1}{V}\sum_{\bfQ} \Gamma_{xy}(\bfQ,q  \hat{z}) \right]. \label{hikamibox2}
\end{eqnarray}
with the weighting factor $\bar{W}^{(f)}_{xy}(\bfq, i\nu_m)$ being given by $\bar{W}^{(e)}_{yx}(-\bfq,i\nu_m)$. Clearly, for $\bfq =0$, it implies 
\begin{eqnarray}
\bar{W}^{(f)}_{xy} = \bar{W}^{(e)}_{yx}=-\bar{W}^{(e)}_{xy}. \label{hikamibox3}
\end{eqnarray} 
Therefore, if the dominant contribution for the CME coefficient is from $\bfq=0$ (to be checked in below), two dressed Hikami boxes, Figs.~\ref{fig1}(e) and (f) cancel each other out. 
In this case, the CME coefficient is determined by $W$ and $\Gamma$. 

Given $W$ and $\Gamma$ with the analytic continuation in frequency $ i \nu_m \rightarrow \omega + i0^+$, 
the CME coefficient can be found by the expanding $ \langle \Pi_{ij}^{{\rm anti}}(q\hat{k}, \omega) \rangle$ ($ij k$ is the cyclic permutation of
$xyz$) in $q$ 
\begin{eqnarray}
\Pi^{{\rm anti}} _{ij}(q\hat{k},\omega)&&\hspace{-1.1em}\approx[W_{ij}^{(0)}(\omega)+qW_{ij}^{(1)}(\omega)][\Gamma_{ij}^{(0)}(\omega)+q\Gamma_{ij}^{(1)}(\omega)]\nonumber \\
&&\hspace{-1.1em}\approx q[W_{ij}^{(0)}(\omega)\Gamma_{ij}^{(1)}(\omega)+W_{ij}^{(1)}(\omega)\Gamma_{ij}^{(0)}(\omega)]
\label{eqexpansion}. 
\end{eqnarray}
Here we have dropped the constant term as they get cancelled in non-superconducting state. $W^{(n)}_{ij}$ and $\Gamma^{(n)}_{ij}$ are coefficients to the $q^n$ term of  $W_{ij}$ and $\frac{1}{V} \sum_Q \Gamma_{ij}$.  Hence the correction of the CME coefficient is obtained as
\begin{eqnarray}
&&\lim_{q\to 0} \Delta\alpha (\mathbf{q},\omega) =  \frac{i}{3} \times \nonumber \\
& & \sum_{(ij)=xy,yz,zx} [W_{ij}^{(0)}(\omega)\Gamma_{ij}^{(1)}(\omega)+W_{ij}^{(1)}(\omega)\Gamma_{ij}^{(0)}(\omega)]. \nonumber \\
\label{deltaalpha}
\end{eqnarray}

From Eq.(\ref{deltaalpha}), it is clear that in addition to the usual term, $\Gamma^{(0)}$, that determines the weak localization/anti-localization behavior of conductivity, the higher order term, $\Gamma^{(1)}$, which is induced self-interacting Cooperon diffusive mode, also contributes the correction of the CME coefficient. As we shall show in below, 
$\Gamma^{(0)}$ and $\Gamma^{(1)}$ propagates in different ways as
\begin{eqnarray}
\hspace{-2em}&&\Gamma_{ij}^{(0)}\thicksim \frac{1}{-i\omega\hspace{-0.2em}+\hspace{-0.2em}Q^2\hspace{-0.2em}+\hspace{-0.2em}1/2\tau_{so}}\label{eq:Gamma0}\\
\hspace{-2em}&&\Gamma_{ij}^{(1)}\thicksim \frac{1}{(-i\omega\hspace{-0.2em}+\hspace{-0.2em}Q^2\hspace{-0.2em}+\hspace{-0.2em}1/2\tau_{so})^2}\label{eq:Gamma1}
\end{eqnarray}
where $\tau_{so}$ is the relaxation time due to the spin-orbit coupling. When the system is clean, one expects $\tau_{so} \rightarrow \infty $ and hence $\Gamma^{(1)} (\omega=0) \sim 1/Q^4$ dominates in Eq.(\ref{deltaalpha}).  As a result, the dominant contribution for the CME coefficient comes from $W_{ij}(\bfq=0,i\nu_m)$ and $\bar{W}_{ij}(\bfq=0,i\nu_m)$. This justifies the conclusion that two dressed Hikami boxes, Figs.~\ref{fig1}(e) and (f) cancel each other out. In addition, it implies that the scaling
of $\delta \alpha$ on the system size $L$ is different from that of the conductivity. As we shall see in below, the CME coefficient exhibits scaling similar to the scaling of localization even in three dimension.

\section{Correction of the CME coefficient}

In this section, we shall find how the CME coefficient scales with the system size $L$ by computing the correction of the CME coefficient due to short-range disorder through calculating $W$ and $\Gamma$.  We first note that the summation over the Matsubara frequency $i \omega_n$ in Eq.(\ref{cooperon}) can be converted into a contour integral over $\varepsilon$ ($i \omega_n \rightarrow \varepsilon$) in the complex plane. After taking the analytic continuation of the external Matsubara frequency $i \nu_n$  to real frequency, $\omega + i 0^+$,  the main contributions come from poles above or below the real axis: $\varepsilon +i 0^+$ or $\varepsilon + i 0^-$, $\varepsilon - \omega + i 0^+$, or $\varepsilon - \omega + i0^- $  \cite{Mahan}. Therefore, near $\omega  \sim 0$, the ladder diagram in Eq.(\ref{cooperon1}) contains combinations of retarded ($G_R$)and advanced ($G_A$) Green's function
in the same frequency: $G_RG_A, G_A G_R, G_R G_R$ and $G_AG_A$.  In the weak scattering limit, after the integration of momentum is done, only $G_RG_A$ and $G_A G_R$ makes significant contribution\cite{Akkermans}. Therefore, we take the analytic continuation
by setting $i\nu_m+i\omega_n \to \omega+\varepsilon+i\eta$ and  $i\omega_n \to \omega-i\eta$ and replacing  $\mathcal{G} (i\omega_n+i\nu_m,\mathbf{k})$ and $\mathcal{G} (i\omega_n,\mathbf{k})$ by $G_R (\omega+\varepsilon,\mathbf{k}) = \frac{1}{\omega+\varepsilon +\epsilon_F-\hat{h}(\mathbf{k})+i\eta}$ and $G_A (\omega,\mathbf{k}) = \frac{1}{\omega+\epsilon_F-\hat{h}(\mathbf{k})-i\eta}$ respectively.  In addition, it is more convenient to express the Green's function in terms of band energies, $E_s$, as
\begin{eqnarray}
G_R(\omega,\mathbf{k}) &=& \frac{\hat{P}_{+}({\bfk})}{\omega +\epsilon_F-E_{+}(\mathbf{k})+i\eta} \nonumber \\
&+& \frac{\hat{P}_{-}({\bfk})}{\omega+\epsilon_F-E_{-}(\mathbf{k})+i\eta}, \label{G+-}
\end{eqnarray}
where $\hat{P}_{s}({\bfk})=\frac{1}{2}(1+s\frac{\mathbf{d}(\mathbf{k})\cdot\bm{\sigma}}{|d(\mathbf{k})|})$ is the projection operator that projects states to two eigen-energies with $s=\pm$\cite{two-band}. 

\subsection{Vertex renormalization and the Drude contribution}
To cope with the contribution from the Cooperon channel, which concerns the $Q \rightarrow 0$ limit, the most important correction
due to the vertex is obtained by taking $\mathbf{q} \rightarrow 0$ in Eq.(\ref{vertex}).  
This is similar to the semi-classical diffusive correction that yields the Drude conductivity, in which the vertex correction, when combined with the non-crossing diagram, leads to the relaxation time correction. Hence, one expects that the vertex simply renormalizes the Fermi velocity in the leading order term\cite{ShenPRB2015}. 

Specifically, in the low energy limit, we set $i\omega_n+i\nu_m=0$. Eq.(\ref{vertex}) becomes 
\begin{eqnarray}
&&\hspace{-1.1em}[\hat{J}_{j}(\bfk\hspace{-0.3em}+\hspace{-0.3em}\bfq)]_{\nu\mu}=[\hat{J}^{0}_{j}(\bfk)]_{\nu\mu}\nonumber\\
&&\hspace{-1.1em}+\frac{1}{V}\sum_{\bfk'}\sum_{\mathcal{C}=\pm}\left[\mathcal{G}^{\mathcal{C}}_{\alpha'\alpha}(\bfk')[\hat{J}_{j}(\bfk\hspace{-0.2em}+\hspace{-0.2em}\bfq)]_{\alpha\beta}\mathcal{G}^{\mathcal{C}}_{\beta\beta'}(\bfk'\hspace{-0.3em}+\hspace{-0.2em}\bfq)\gamma_{e}b^{\alpha'\mu}_{\beta'\nu}\right]\nonumber \\
&&\hspace{-1.1em}+\frac{1}{V}\sum_{\bfk'}\sum_{\bar{\mathcal{C}}, \mathcal{C}=\pm}\left[\mathcal{G}^{\bar{\mathcal{C}}}_{\alpha'\alpha}(\bfk')[\hat{J}_{j}(\bfk\hspace{-0.2em}+\hspace{-0.2em}\bfq)]_{\alpha\beta}\mathcal{G}^{\mathcal{C}}_{\beta\beta'}(\bfk'\hspace{-0.3em}+\hspace{-0.2em}\bfq)\gamma_{e}b^{\alpha'\mu}_{\beta'\nu}\right] . \nonumber \\ \label{vertex correction} 
\end{eqnarray}
Here $\bar{\mathcal{C}}$ and $\mathcal{C}$ are the indices that denote Weyl nodes $\mathcal{C}^{\pm}$. We have separated the product $\mathcal{G}\mathcal{G}$ as contributions by inter-nodes ($\mathcal{G}^{\bar{\mathcal{C}}}\mathcal{G}^{\mathcal{C}}$)and intra-nodes ($\mathcal{G}^{\mathcal{C}}\mathcal{G}^{\mathcal{C}}$). 
By taking $\mathbf{q}=0$ in Eq.(\ref{vertex correction}), we obtain
\begin{eqnarray}
\hat{J}_{j}(1-\gamma_e \check{\Lambda} b)=\hat{J}^{0}_{j}, \nonumber \\
\end{eqnarray}
where $b$ is the corresponding matrix to $b^{\alpha'\mu}_{\beta'\nu}$ that describes the correlation of disorder and is a $4\times 4$ unit matrix for spin-independent disorders.  $\check{\Lambda}$ is the particle-hole propagator. 
Since the total momentum of inter-node Fermions does not vanish, the inter-node contribution drops and hence $\check{\Lambda}$ is given by
\begin{eqnarray}
\check{\Lambda}^{\alpha\alpha'}_{\beta\beta'}=\frac{1}{V}\sum_{\bfk}\sum_{\mathcal{C}=\pm}(\mathcal{G}^{\mathcal{C}}_{\alpha'\alpha}\otimes\mathcal{G}^{\mathcal{C}}_{\beta\beta'}) 
\end{eqnarray}
Replacing $\mathcal{G}$ by $G_A$ or $G_R$,  $\check{\Lambda}$ for intra-Weyl node ($\mathcal{C}_{+}$) can be written as 
\begin{eqnarray}
\hspace{-2.5em}\bar{\Lambda}&&\hspace{-1em}=\frac{1}{V}\sum_{\bfk} G^T_R\hspace{-0.2em}\otimes\hspace{-0.2em}G_A\nonumber \\
&&\hspace{-1em}=\frac{1}{4\pi}\int d\Omega\int^{\infty}_{0}|\bfk|^2dk\sum_{s,t=\pm}\frac{1}{\omega +\nu \hspace{-0.2em}+\hspace{-0.2em}\mu\hspace{-0.2em}-\hspace{-0.2em}E_{s}\hspace{-0.2em}+\hspace{-0.2em}i\eta}\nonumber \\
&&\hspace{3em}\times\frac{1}{\omega \hspace{-0.2em}+\hspace{-0.2em}\mu\hspace{-0.2em}-\hspace{-0.2em}E_{s}\hspace{-0.2em}-\hspace{-0.2em}i\eta}\hat{P}^T_{s}({\bfk})\otimes\hat{P}_{s}({\bfk}), \nonumber \\ \label{Lambda1}
\end{eqnarray}
where $\hat{P}_{s}({\bfk})=\frac{1}{2}(1+s\frac{\mathbf{k}\cdot\bm{\sigma}}{|k|})$ is the projection operator for the Weyl node and $s$ is the band index with eigen-energies $E_{\pm}$. Following Refs.[\onlinecite{Akkermans,MacDonald2014}], the integration over the magnitude $k$ is replaced by integrations over the energy $\epsilon$, $\int^{\Lambda_1}_{-\Lambda_2} d \epsilon N(\epsilon_F)$, which can be then evaluated by complex contour integrals. For $\mathbf{k}$ being close to $(0,0,0)$, one takes $s=-$; while for $\mathbf{k}$ being close to $(0,0,\pi)$,
one takes one takes $s=+$. By setting $\omega=0$ and taking the analytic continuation of the integration of $k$ in the
complex plane and performing the angular integration, we find 
\begin{eqnarray}
\check{\Lambda}&&\hspace{-1em}=-\frac{2\pi N(\epsilon_{F})\tau_e}{6} \begin{pmatrix}                
2 & 0 & 0 & 1 \\
0 & 1 & 0 & 0 \\
0 & 0 & 1 & 0 \\
1 & 0 & 0 & 2
\end{pmatrix}. \nonumber \\ \label{Lambda2}
\end{eqnarray}
Similar analysis for the other Weyl node yields the same result.  By using the identity, $2\pi N(\epsilon_{F})\tau_e =1/ \gamma_e$, the vertex correction is given by
\begin{eqnarray}
\hat{J}_{j}
\begin{pmatrix}                
\frac{4}{3} & 0 & 0 & \frac{1}{6} \\
0 & \frac{7}{6} & 0 & 0 \\
0 & 0 & \frac{7}{6} & 0 \\
\frac{1}{6} & 0 & 0 & \frac{4}{3}
\end{pmatrix}
=\hat{J}^{0}_{j}, \label{eq:vertex on current}\nonumber \\
\end{eqnarray}
where $\hat{J}_{j}$ and $\hat{J}^{0}_{j}$ are arranged into vectors with 4 components. Note that if we rearrange 
$\hat{J}_{j}$ and $\hat{J}^{0}_{j}$
back to $2 \times 2$ matrices, Eq.(\ref{eq:vertex on current}) implies $\hat{J}_{j} ({\bf k}) = \frac{6}{7} \hat{J}^{0}_{j} ({\bfk})$.

\begin{figure}[t]
\includegraphics[height=2.5in,width=3.2in] {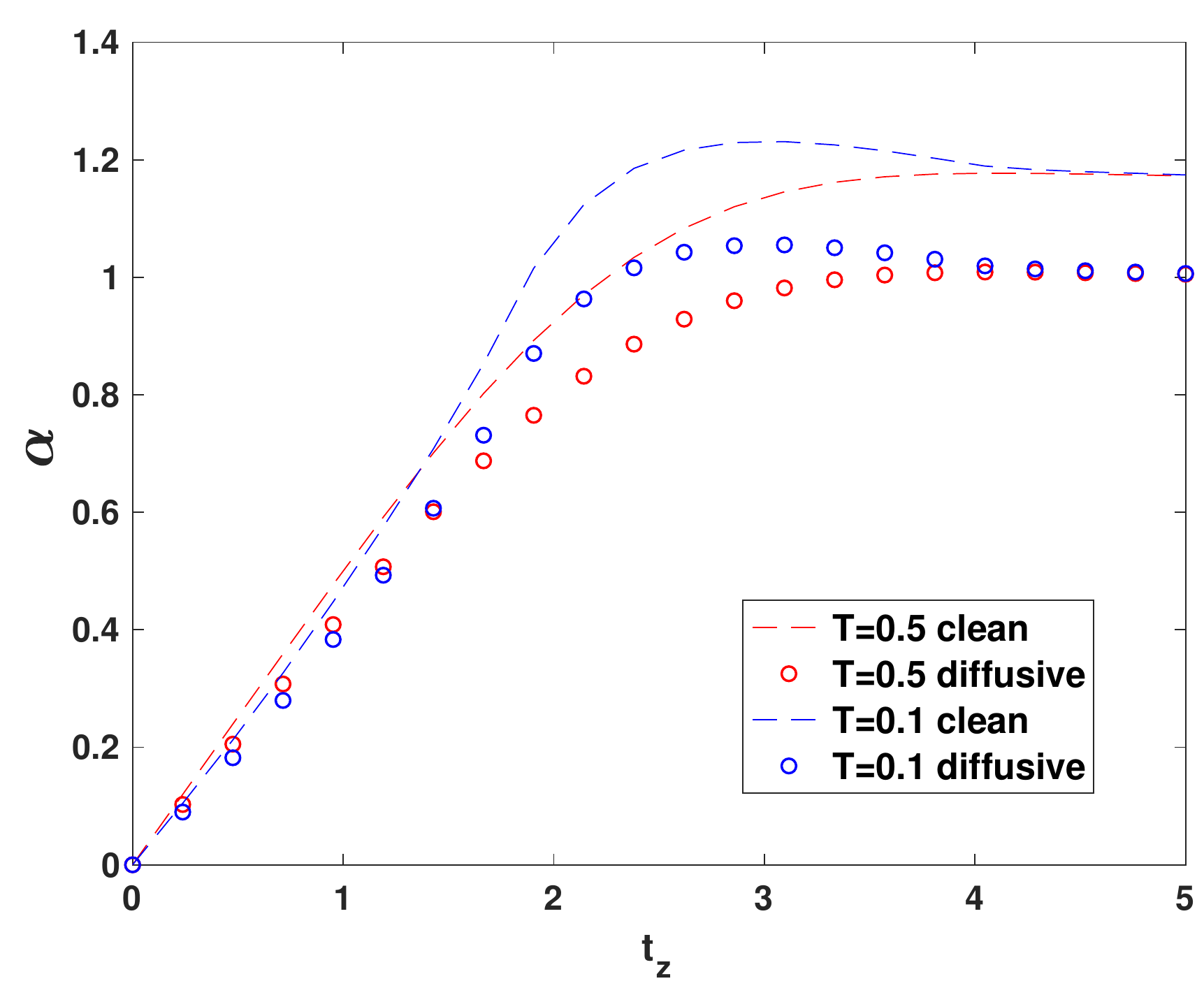} 
\caption{Rernormalized total CME coefficient due to the Drude correction. Here dash lines reproduce the CME coefficient $\alpha$ of Ref.[\onlinecite{two-band}] in the clean limit.  Open circles are the renormalized total CME coefficient with the Drude correction.}
\label{fig2}
\end{figure}
The above vertex renormalization, when combined with the non-crossing diagram, gives rise to the
total CME coefficient due to the Drude correction.  Since $\hat{J}_{j} ({\bf k}) = \frac{6}{7} \hat{J}^{0}_{j} ({\bfk})$, 
the CME coefficient is simply renormalized by a factor $7/6$ after including the Drude correction.
In Fig.~1, we show the total CME coefficient due to the Drude correction. In comparion to the clean limit given by Ref.[\onlinecite{two-band}],
$\alpha$ gets renormalized by a factor $7/6$ due to disorders in the Drude correction. However, the Drude correction is size-independent and and the total CME coefficient remains finite at the renormalized value.  In the following,
we will show that for systems with larger sizes $L$, the CME coefficient will get further suppressed by the Cooperon channel.

\subsection{Cooperon Propagator}
The Cooperon propagator, $\Gamma$, can be found by solving Eq.(\ref{ladder diagram}), which can be rewritten in a matrix form as
\begin{eqnarray}
\Gamma\left[\gamma_{e}^{-1}b^{-1}-\frac{1}{V}\sum_{\bfk"}\mathcal{G}\otimes\mathcal{G}\right]=1.
\end{eqnarray}\\
Hence $\Gamma^{-1} = \gamma_{e}^{-1}b^{-1}-\frac{1}{V}\sum_{\bfk"}\mathcal{G}\otimes\mathcal{G}$.
To find the leading orders of $\Gamma$, $\Gamma^{(0)}$ and $\Gamma^{(1)}$, we shall expand $\Gamma^{-1}$ with respect
to $\mathbf{q}$ in different directions and keep the order of $q$. The component of $\Gamma_{ij}^{-1}$ corresponding to
the direction $\hat{k}$ ($\mathbf{q}=q_k \hat{k}$)with $i$, $j$ and $k$ being cyclic permutations of $x$, $y$, and $z$ is given by 
\begin{equation}
\Gamma_{ij}^{-1}(\bfQ)= A_{ij} (\bfQ)+ q_k B_{ij} (\bfQ). \label{gamma-1}
\end{equation}
Following Ref.[\onlinecite{MacDonald2014}], by symmetrizing $\bfQ+\bfq$ between two Green's function in Eq.(\ref{ladder diagram}), expanding
the wave-vector $\mathbf{k} \pm (\bfQ+\bfq)$ with respect to $\mathbf{k}$ to leading orders ($O(Q^2)$), and replacing $\mathcal{G}$ by
$G_R$ or $G_A$ , we obtain
\begin{eqnarray}
P &\equiv&\frac{1}{V}\sum_{\bfk} G_R (\bfk\hspace{-0.2em}+\hspace{-0.2em}\frac{\tilde{\bfQ}}{2}, i\omega_n\hspace{-0.2em}+\hspace{-0.2em}i\nu_m)\otimes G_A(-\bfk\hspace{-0.2em}+\hspace{-0.2em}\frac{\tilde{\bfQ}}{2},i\omega_n)\nonumber \\
&&\hspace{-0.5em}=\frac{1}{V}\sum_{\bfk}\frac{1}{G_R^{-1} (\bfk)-\Lambda}\otimes\frac{1}{G_A^{-1}(-\bfk)-\Lambda}\nonumber \\
&&\hspace{-0.5em}\approx\frac{1}{V}\sum_{\bfk} (G_R+G_R\Lambda G_R+G_R \Lambda G_R \Lambda G_R) \nonumber \\
&&\hspace{3em}\otimes (G_A+G_A\Lambda G_A+G_A \Lambda G_A \Lambda G_A),\nonumber \\
\end{eqnarray}
where $G_{R/A}= G ( \omega, \pm \bfk)$, $\tilde{\bfQ}=\bfQ+\bfq$ and $\Lambda=\frac{v_{F}}{2} [\tilde{\bfQ} \cdot\bm{\sigma}] $. Hence $P$ is a summation of three contributions $P=P^{(0)}+P^{(1)}+P^{(2)}$ with
\begin{eqnarray}
&& P^{(0)}=\frac{1}{V}\sum_{\bfk}(G_R\hspace{-0.2em}\otimes\hspace{-0.2em}G_A),\nonumber \\
&&P^{(1)}=\frac{1}{V}\sum_{\bfk}\left[(G_R \Lambda G_R)\hspace{-0.2em}\otimes G_A\hspace{-0.2em}+\hspace{-0.2em}G_R\hspace{-0.2em}\otimes\hspace{-0.2em}(G_A\Lambda G_A)\right],\nonumber \\
&&P^{(2)}=\frac{1}{V}\sum_{\bfk}\left[(G_R\Lambda G_R\Lambda G_R)\hspace{-0.2em}\otimes\hspace{-0.2em}G_A\hspace{-0.2em} \right. \nonumber \\
&&\left. + G_R\otimes(G_A \Lambda G_A \Lambda G_A)
+G_R\Lambda G_R\otimes\hspace{-0.2em}G_A \Lambda G_A \right].\nonumber \\
\end{eqnarray}
To find $P^{(n)}$, we note that $P$ is invariant under rotations. Hence $P^{(n)}$ must
be a combination of $\tilde{\bf Q}^2$, $\tilde{\bf Q} \cdot \bf{S}_d$, and $\tilde{\bf Q} \cdot \bf{S}$ with $\bf{S}_d = \bm{\sigma} \otimes \bm{1} - \bm{1} \otimes \bm{\sigma}$  and $\bf{S} = \bm{\sigma} \otimes \bm{1} + \bm{1} \otimes \bm{\sigma}$ being the difference of spin and the total spin operators of the Cooperon respectively.
Consider the Weyl node ${\mathcal{C}}_{+}$, by performing similar integrations as what were done in Eqs.(\ref{Lambda1}) and (\ref{Lambda2}), we find 
\begin{eqnarray}
\hspace{-2.5em}P^{(0)} =- \frac{1}{6\gamma_e} \begin{pmatrix} 
1& 0 & 0 & 0 \\
0 & 2 & -1& 0 \\
0 & -1 & 2& 0 \\
0 & 0 & 0 & 1               
\end{pmatrix},\nonumber \\
\end{eqnarray}
while  we find that
\begin{eqnarray}
P^{(1)} = -\frac{iv_F \tau_e}{12 \gamma_e}  
\tilde{\bf Q} \cdot \bf{S}_d,   \label{P1}
\end{eqnarray}
and 
\begin{eqnarray}
P^{(2)} = \frac{v^2_F \tau^2_e}{60 \gamma_e} \left[  - 20 \tilde{Q}^2 \bm{1}+ \tilde{Q}^2 \bm{S}^2 +2 (
\tilde{\bf Q} \cdot \bf{S})^2 \right].  \label{P2}
\end{eqnarray}

$A_{ij} (\bfQ)$ in Eq.(\ref{gamma-1}) is then obtained by setting $\tilde{\bf Q} = {\bf Q}$ and summing over
$P^{(0)}$,  $P^{(1)}$, and $P^{(2)}$. To obtain
$B_{ij}(\bfQ)$,  we set $\tilde{\bf Q} = {\bf Q} + q \hat{k}$ in $P^{(n)}$ and keep $O(q)$ terms.
For $ij=xy$ and $\mathbf{q} = q \hat{z}$, we find
\begin{eqnarray}
B_{xy}(\bfQ) &=& \frac{v^2_F \tau^2_e}{30 \gamma_e} \left[  -20 Q_z \bm{1}+ Q_z \bm{S}^2 +\right. \nonumber \\
&& \left. ({\bf Q} \cdot {\bf S}) S_z +S_z  (
{\bf Q} \cdot {\bf S})\right]. 
\end{eqnarray}

To obtain $\Gamma_{ij}$, it is more convenient to
use  the singlet-triplet basis of the total spin,  $\{\ket{S=1, S_z =1},\ket{S=1, S_z=0,},\ket{S=1, S_z=-1},\ket{S=0} \}$ by performing a similar transformation on $A_{ij}(\bfQ)$ and $B_{ij}(\bfQ)$ with
\begin{eqnarray}
\begin{pmatrix}
 \ket{1,\hspace{0.67em}1}\\
 \ket{1,\hspace{0.67em}0}\\
 \ket{1,-1}\\
 \ket{0,\hspace{0.67em}0} 
\end{pmatrix}=  
\begin{pmatrix}   
1& 0 & 0& 0 \\
0  & \frac{1}{\sqrt{2}}  &  \frac{1}{\sqrt{2}} &0\\
0 & 0 & 0 & 1 \\
0 & \frac{1}{\sqrt{2}} & -\frac{1}{\sqrt{2}} & 0 
\end{pmatrix}
\begin{pmatrix}
 \ket{\uparrow\uparrow}\\
 \ket{\uparrow\downarrow}\\
 \ket{\downarrow\uparrow}\\
 \ket{\downarrow\downarrow}
\end{pmatrix}
\equiv S 
\begin{pmatrix}
 \ket{\uparrow\uparrow}\\
 \ket{\uparrow\downarrow}\\
 \ket{\downarrow\uparrow}\\
 \ket{\downarrow\downarrow}
\end{pmatrix}
\nonumber\\
\end{eqnarray}
Under this basis, $A_{xy}(\bfQ)$ and $B_{xy}(\bfQ)$ (replaced by $S^{-1}A_{xy}(\bfQ)S$
and $S^{-1}B_{xy}(\bfQ)S$) are given by
\begin{eqnarray}
A_{xy}(\bfQ) &=& \frac{2}{\gamma_e} \bm{1} - \frac{2}{3\gamma_e} \bm{1}_3  + i \frac{\lambda_2}{12}
{\bf Q} \cdot \bf{S}_d\nonumber \\
&+& \frac{\lambda_3}{20} \left[ 20Q^2 \bm{1}- 8 Q^2 \bm{1}_3 -2 (
{\bf Q} \cdot \bf{S})^2 \right]
\end{eqnarray}
and
\begin{eqnarray}
B_{xy}(\bfQ) &=&   i \frac{\lambda_2}{12} S^z_d + \frac{\lambda_3}{10 } \left[  20 Q_z \bm{1} - 8 Q_z \bm{1}_3 \right. \nonumber \\
&-& \left. ({\bf Q} \cdot {\bf S}) S_z -S_z  (
\bf Q \cdot \bf{S})\right],
\end{eqnarray}
where $\lambda_2 =  v_F \tau_e /\gamma_e$, $\lambda_{3}=\tau_e D / \gamma_e$ with $D=v_{F}^{2}\tau_e /3$ being the diffusion constant and relevant matrices are given by
\begin{eqnarray}
&& \bm{1}_3 =  \begin{pmatrix}   
1&0 & 0 & 0 \\
 0 &1  &0  & 0 \\
0 &  0 & 1  & 0 \\
0 & 0 & 0 & 0
\end{pmatrix},  \nonumber \\
{\bf Q} \cdot \bf{S}_d  &=&\nonumber \\
&& \begin{pmatrix}   
0 &  0 & 0 & -\sqrt{2} Q_{-}  \\
 0 & 0  &  0 & 2 Q_z \\
0 &  0 & 0  & \sqrt{2} Q_{+} \\
 -\sqrt{2} Q_{+} & 2 Q_z & \sqrt{2} Q_{-} & 0
\end{pmatrix}, \nonumber \\
\end{eqnarray}
and
\begin{eqnarray}
{\bf Q} \cdot \bf{S}  &=&
\begin{pmatrix}   
2 Q_z &  \sqrt{2} Q_{-} & 0 & 0  \\
 \sqrt{2} Q_{+} & 0  &  \sqrt{2} Q_{-} & 0 \\
0 &  \sqrt{2} Q_{+} & -2 Q_z  & 0 \\
 0 & 0 & 0 & 0
\end{pmatrix}, \nonumber \\
\end{eqnarray}
with $Q_{\pm} =  Q_x \pm i Q_y$.
Similarly, we find 
$A_{yz}(\bfQ) = A_{zx}(\bfQ) = A_{xy}(\bfQ)$, while for $B_{yz}$ and $B_{zx}$, we find that ($Q_z$, $S_z$) in  $B_{xy}$ is replaced by ($Q_x$, $S_x$) and ($Q_y$, $S_y$) respectively.

The total intra-node Cooperon propagator for the node $\mathcal{C}^+$ can be obtained from Eq.(\ref{gamma-1}) as $\Gamma(\bfQ)\simeq A(\bfQ)^{-1}-qA(\bfQ)^{-1}B(\bfQ)A(\bfQ)^{-1}$.  After performing integration over angles and keeping terms up to $O(Q^2)$, we find
\begin{eqnarray}
\hspace{-2em}&&\Gamma^{(0)}_{xy}(Q)=\begin{pmatrix}                
\Gamma^{(0)\alpha}_T & 0 & 0 & 0 \\
0 & \Gamma^{(0)\beta}_T & 0 & 0 \\
0 & 0 & \Gamma^{(0)\alpha}_T & 0 \\
0 & 0 & 0 & \Gamma^{(0)}_S 
\end{pmatrix},\label{Gamma0}
\end{eqnarray}
where the coefficients are given by
\begin{eqnarray}
&& \Gamma^{(0)\alpha}_T=\frac{ \pi \gamma_e}{\lambda^2}\left[ \frac{\alpha}{Q^2+\lambda'^{-2}}+\frac{\beta}{Q^2+\lambda^{-2}}\right],\nonumber \\
&& \Gamma^{(0)\beta}_T=\frac{3\pi \gamma_e}{4 \lambda^2} \frac{1}{Q^2+\lambda^{-2}},\nonumber \\
&&\Gamma^{(0)}_S=\frac{ \pi \gamma_e}{2 \lambda^2}\frac{1}{Q^2+\lambda^{-2}}.  \label{Gamma01}
\end{eqnarray}
Here $\alpha \approx 9.29$ and $\beta \approx 4.21$. 
 $\lambda^{-2} =7680/[719 ( v_F \tau_e)^2]$ and $\lambda'^{-2} = (719/384)\lambda^{-2}$ are the induced gaps due to the spin-orbit interaction. $\lambda$ is the characteristic spin-relaxation length scale for the propagation of the Cooperon. 

Similarly, for other components, the equality, $A_{yz}(\bfQ) = A_{zx}(\bfQ) = A_{xy}(\bfQ)$, implies that
$\Gamma^{(0)}_{yz}(Q)=\Gamma^{(0)}_{zx}(Q)=\Gamma^{(0)}_{xy}(Q)$.
On the other hand, after performing integration over angles and keeping terms up to $O(Q^2)$, we find 
\begin{eqnarray}
&&\hspace{-3em}\Gamma^{(1)}_{xy}(Q)= \Gamma^{(1)}_{xy,T}
\begin{pmatrix}   
 0 &  0 & 0 & 0 \\
0 & 0  &  0 & i \\
0 & 0 & 0 & 0 \\
0 & i  & 0 & 0  
\end{pmatrix}, \label{Gamma1_1}
\end{eqnarray}
where the coefficient is given by
\begin{eqnarray}
\Gamma^{(1)}_{xy,T}=-\sqrt{\frac{15}{1438}}\frac{ \pi  \gamma_e}{4 \lambda^3}\frac{1}{(Q^2+\lambda^{-2})^2}. \nonumber \\ \hspace{-10em}
\end{eqnarray}
Similar calculations show that
\begin{eqnarray}
&&\hspace{-3em}\Gamma^{(1)}_{yz}(Q)= \Gamma^{(1)}_{yz,T}
\begin{pmatrix}   
 0 &  0& 0 & i \\
0 & 0  &  0& 0 \\
0 & 0 & 0 & -i \\
i & 0 & -i & 0  
\end{pmatrix},  \nonumber \\
&&\hspace{-3em}\Gamma^{(1)}_{zx}(Q)= \Gamma^{(1)}_{zx,T}
\begin{pmatrix}   
 0 &  0 & 0 & 1 \\
0 & 0  &  0 & 0\\
0 & 0 & 0 & 1 \\
-1 & 0 & -1 & 0  
\end{pmatrix}, \label{Gamma1}
\end{eqnarray}
where the coefficients are given by
\begin{eqnarray}
&& \Gamma^{(1)}_{yz,T}= \Gamma^{(1)}_{zx,T} \equiv   \Gamma^{(1)}_{T} = \sqrt{\frac{15}{719}}\frac{ \pi  \gamma_e}{ 8 \lambda^3}\frac{1}{(Q^2+\lambda^{-2})^2} \nonumber \\  \label{Gamma11}
\end{eqnarray}
Similar calculation can be carried out for the Weyl node ${\mathcal{C}}^{-}$. Due to its different helicity, we find that $\Gamma^{(0)}(Q)$ is the same, while $\Gamma^{(1)}_{xy}(Q)$ is opposite in sign. As a result, $\Gamma^{(1)}_{xy}(Q)$ gets cancelled. 

Finally, because the total momentum of  two electrons near the same Weyl node $\mathcal{C}^{\pm}$ is close to $Q \sim 0$ (or $2\pi$), the intra-node Cooperon is dominated by $Q \sim 0$ as described in the above. However, for the Cooperon channel composed by inter-node Fermions., the total momentum is around $\mathbf{Q} \sim (0,0,\pi)$. Hence  there is no contribution for $Q \sim 0$. The contribution due to Cooperon channel is thus determined by Eq.(\ref{deltaalpha}) with $\Gamma^{(0)}$ being given by Eqs.(\ref{Gamma0}) and (\ref{Gamma01}) and $\Gamma^{(1)}$ being given by Eqs.(\ref{Gamma1}) and (\ref{Gamma11}).

\section{Numerical Results}
From Eq.(\ref{deltaalpha}), the  correction to the CME coefficient due to the Cooperon channel is determined by $\Gamma^{(0)}$ and $\Gamma^{(1)}$.  It is clear from Eqs. (\ref{Gamma0}), (\ref{Gamma01}), (\ref{Gamma1}) and (\ref{Gamma11}) that in the limit $Q \rightarrow 0$,
$\Gamma^{(0)} \sim 1/Q^2$ while $\Gamma^{(1)} \sim 1/Q^4 $.  Since the correction of the Cooperon channel is proportional to the integral over
$\mathbf{Q}$, the dominant contribution is given by
\begin{eqnarray}
&& \Delta\alpha=\frac{2i}{3}  \int \frac{Q^2 dQ }{(2\pi)^3}  \Tr \left[ W^{(0)}_{yz} \Gamma^{(1)}_{yz} +W^{(0)}_{zx} \Gamma^{(1)}_{zx}  \right], \nonumber \\
\label{Q4}
\end{eqnarray}  
where the weighting factors  $W^{(0)}_{ij}$ are derived in Appendix A and need to be computed numerically
by using Eqs. (\ref{GGGG}) and (\ref{Wcomp}).

The contribution of the propagator, $ \frac{1}{(Q^2+\lambda^{-2})^2}$, to the CME coefficient results in different scaling behaviors from those for conductivity. Similar to the scaling analysis done on the conductivity, it is useful to analyze the change of the CME coefficient versus the system size by defining the $\beta$ function as
\begin{eqnarray}
\beta_{{\rm CME}} = \frac{d \ln |\Delta \alpha (L)|}{d \ln L}. \label{betaf}
\end{eqnarray}
The $\beta$ function is the characteristic function that determines how the CME coefficient scales with the system size. When $\alpha \Delta \alpha <0$ (i.e. for positive $\alpha$, $\Delta \alpha<0$), if the $\beta$ function is positive (i.e., $d \Delta \alpha /d L <0$), the quantum correction exhibits localization behavior and the CME is suppressed in bulk Weyl semimetals. On the other hand, when $\alpha \Delta \alpha >0$ (i.e. for positive $\alpha$, $\Delta \alpha>0$),  if the $\beta$ function is positive (i.e., $d \Delta \alpha /d L >0$), the quantum correction exhibits anti-localization behavior and the CME is enhanced in bulk Weyl semimetals.

To find the scaling behavior of the CME coefficent, it is important to know that there are three characteristic length scales that determine the quantum correction of the CME coefficient: mean free path $l$, the system size $L$ or phase coherent length, and the spin-relaxation length $\lambda$ as
\begin{eqnarray}
\Delta \alpha &\propto& \int ^{l^{-1}}_{L^{-1}}\hspace{-0.2em}dQ Q^2 \frac{1}{(Q^2+\lambda^{-2})^2}\hspace{-0.2em}\nonumber  \\
&=&\hspace{-0.2em}\left\{ \frac{L}{2[1+(L/\lambda)^2]}- \frac{l}{2[1+(l/\lambda)^2]} \right. \nonumber \\
&& \left. +\frac{\lambda}{2} \left[ \tan^{-1} (\frac{\lambda}{l})-\tan^{-1}\hspace{-0.1em}(\frac{\lambda}{L}) \right] \right\}.\label{1/Q^4}
\end{eqnarray}
\begin{figure}[t]
\includegraphics[width=7.2cm,height=5.6cm]{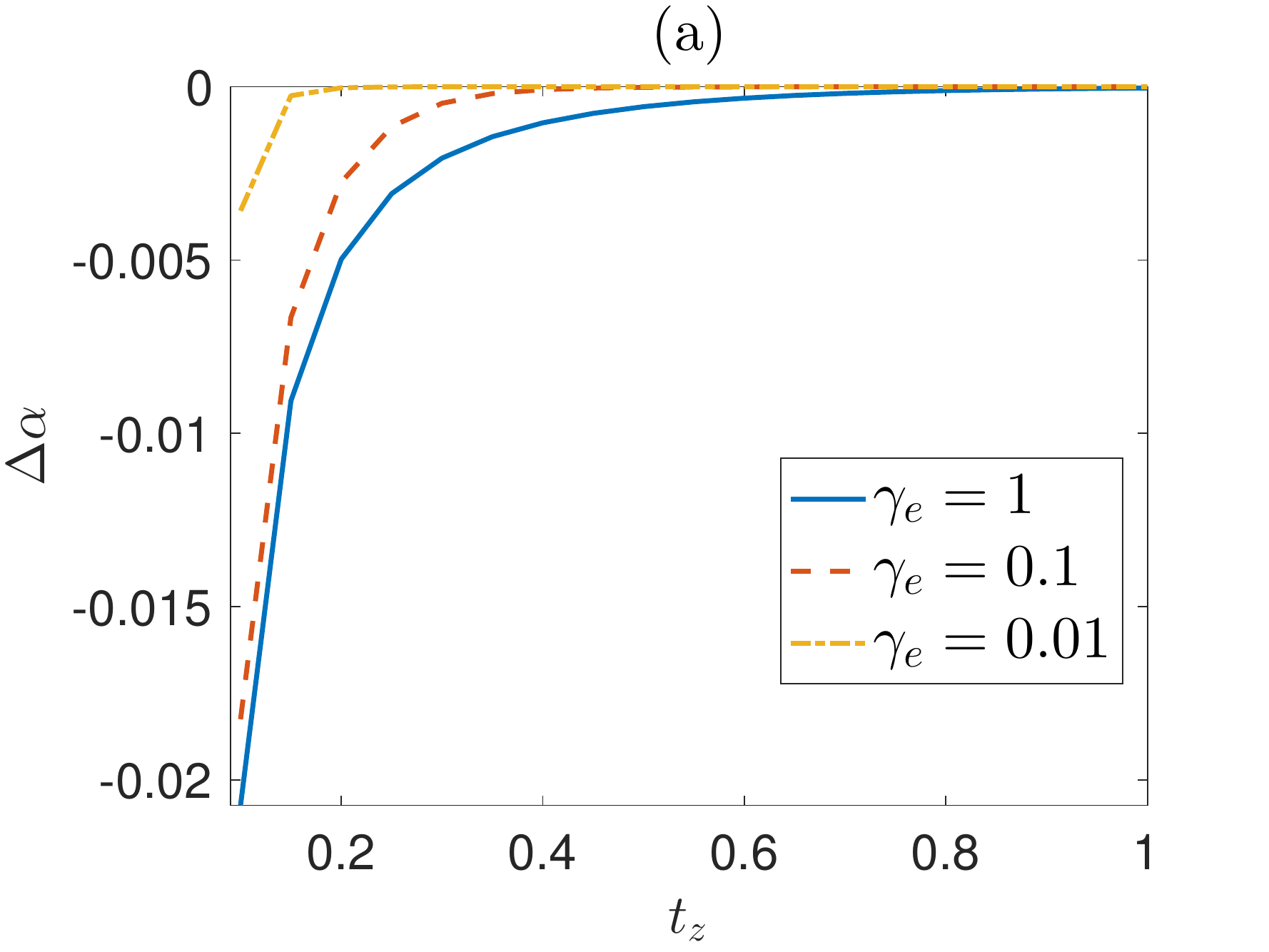} 
\includegraphics[width=6.8cm,height=5.4cm]{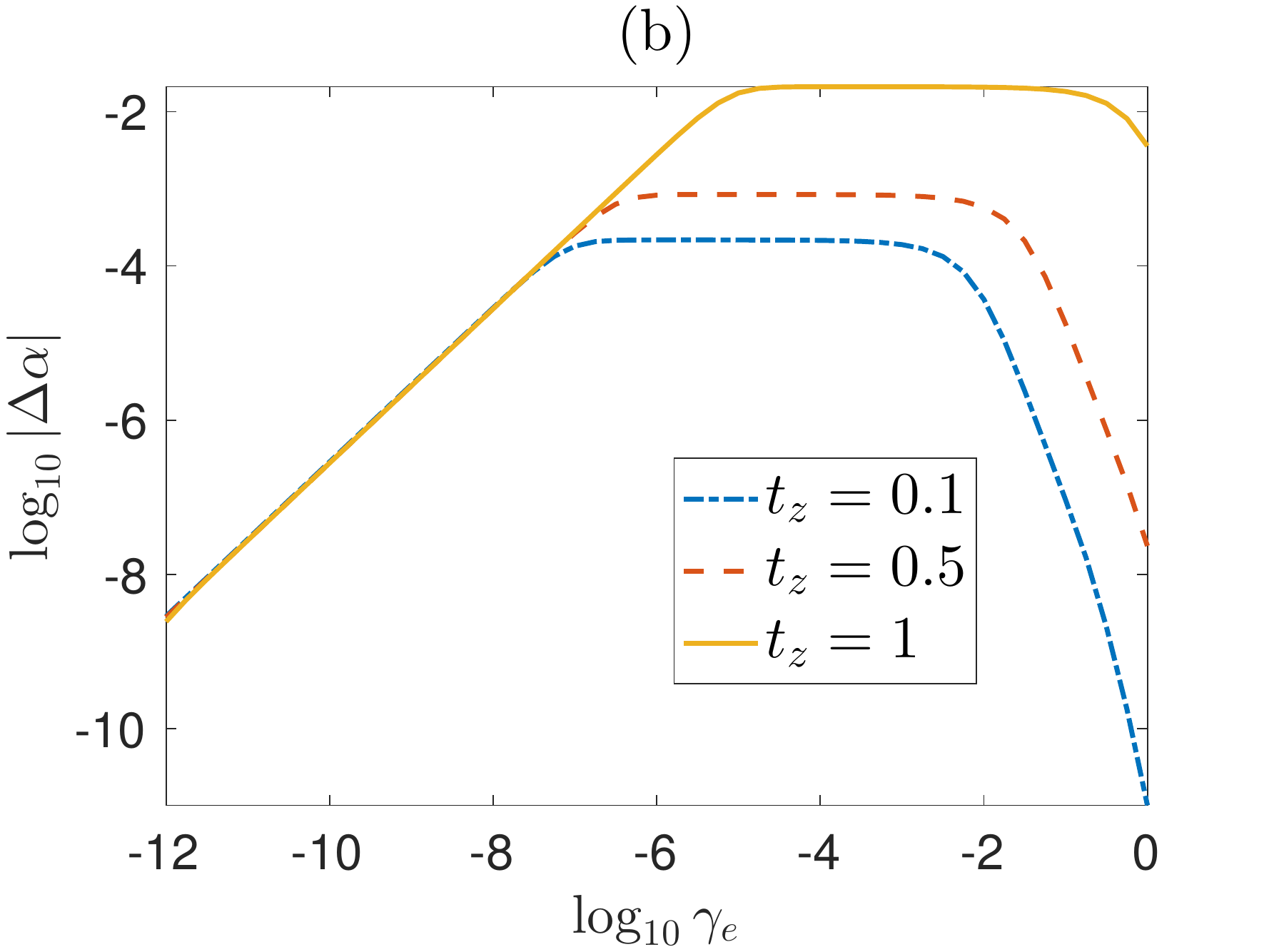} 
\caption{Correction of  the CME coefficient due to the Cooperon channel, 
$\Delta \alpha$ (in unit of $\frac{e^2 \lambda_{so}}{\hbar}$), due to short-range disorder.
 (a) $\sgn (\alpha)$ $\Delta \alpha$ versus $t_z$ for different  $\gamma_e$ in the intermediate region
 $l<\lambda< L$.  (b) $\log_{10} |\Delta \alpha|$ versus disorder strength $\gamma_e$ for different $t_z$. Here $\lambda_{so}=1$, $T=0.001$, $\epsilon_F=0$, $l=100$, and the system size $L/a =10^7$ with $a$ being the lattice constant.
\label{delta_alpha}}
\end{figure}
In the clean limit with  $l<L\ll \lambda$, i.e., the spin-relaxation length is larger than system size, the contribution of Cooperon diverges due to the the infrared divergence in the integration of $1/Q^4$ terms.  The dominating quantum interference is proportional to $L$, which shows that the quantum correction is linearly proportional to the effective system size. Hence the quantum correction due to the Cooperon channel eventually wins over the Drude contribution.
 On the other hand, when $l<\lambda< L$, the spin-relaxation due to impurities can no longer be neglected. In this intermediate region, the integration over $Q$ is finite, leading to a finite correction to the CME coefficient. 
Finally, when the spin-relaxation length is shorter than the mean free path, $\lambda\lesssim l<L$, the quantum correction is absent. 
\begin{figure}[t]
\includegraphics[width=6.5cm,height=5.5cm]{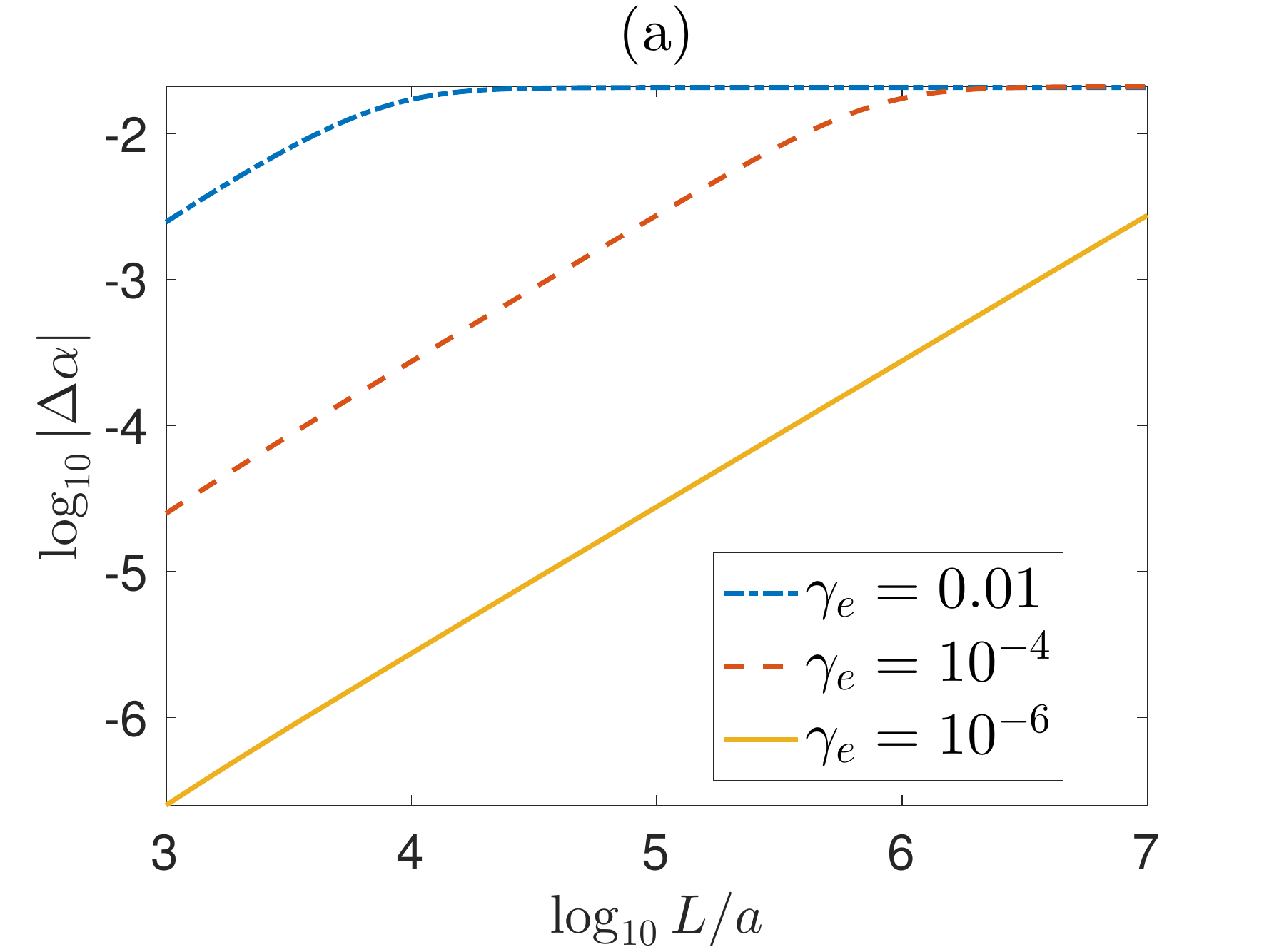} 
\includegraphics[width=6.6cm,height=5.5cm]{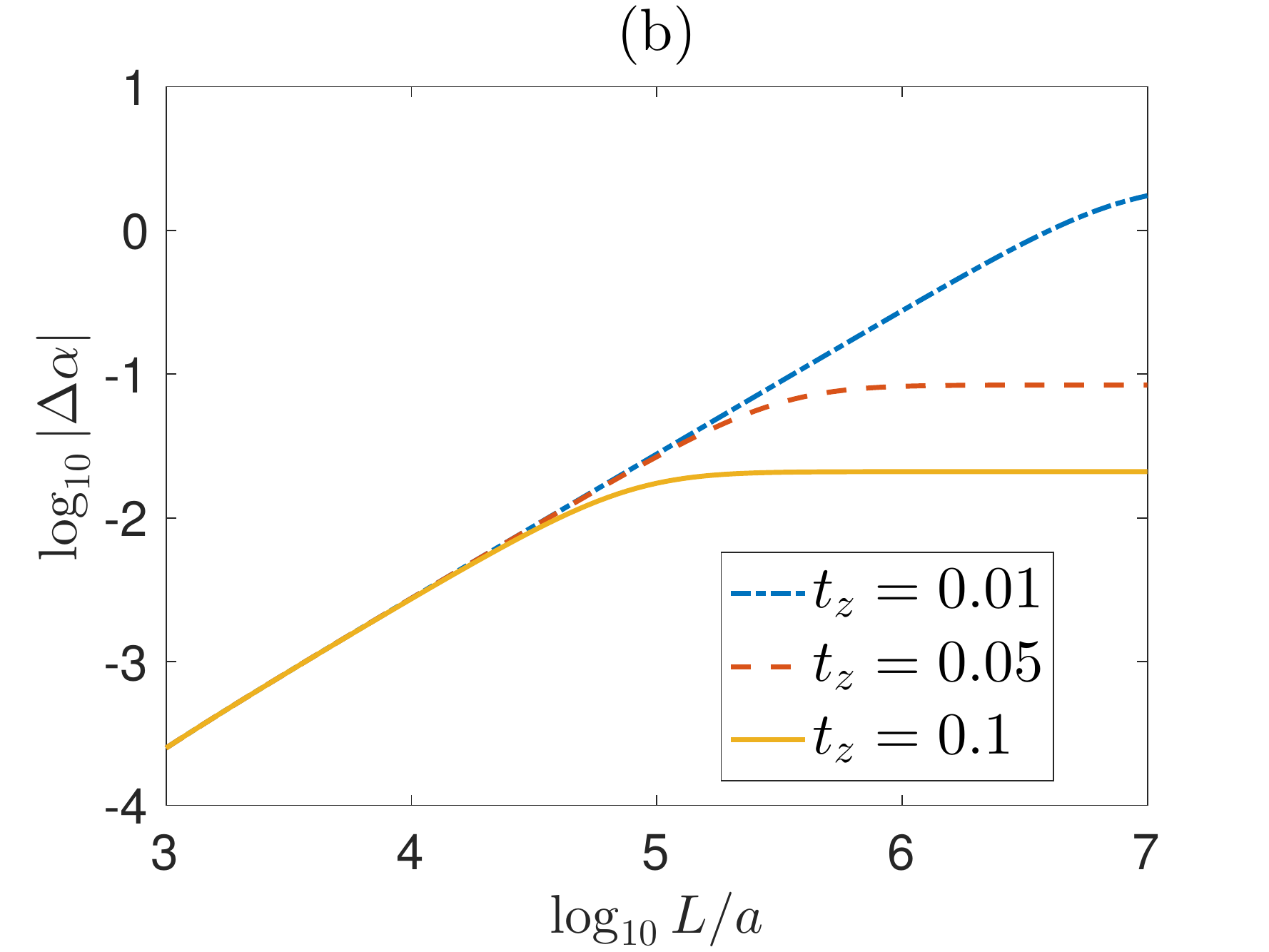}
\caption{(a) Transition behavior of $\log_{10} |\Delta \alpha|$ from the intermediate region, $l<\lambda< L$, to the clean limit $l<L\ll \lambda$. Here  $t_z=0.1$, $\lambda_{so}=1$, $T=0.001$, $\epsilon_F=0$, and $l=100$. The clean limit, $\gamma_e=10^{-6}$, exhibits the scaling behavior with $\Delta \alpha \propto -(L/a)$. (b) $\log_{10} |\Delta \alpha|$ versus $\log L/a$ for different  $t_z$. It is seen that $\beta_{{\rm CME}}$ is always positive and approaches zero for larger systems. Here $\gamma_e=0.001$, $\lambda_{so}=1$, $T=0.001$, $\epsilon_F=0$, and $l=100$.
\label{transition}} 
\end{figure}

By including the weighting factor and the summation over 4 Green's fucntions as derived in Appendix A, the correction of the CME coefficient  is computed numerically.   
In Fig.~\ref{delta_alpha}(a), we show $\sgn (\alpha) \Delta \alpha$ versus $t_z$ for $k_B T=0.001$ with different scattering  strengths in the intermediate region $l<\lambda< L$. Since $\alpha>0$,
it is seen that $\Delta \alpha$ is always negative, indicating the suppression of the CME coefficient due to short-range disorder.  Fig.~\ref{delta_alpha}(b) shows how $\log |\Delta \alpha|$ depends on the disorder strength $\gamma_e$ for different $t_z$.
In the intermediate region $l<\lambda< L$ (larger $\gamma_e$), the scale factor of $\Delta \alpha$ increases with $\lambda$ as indicated in Eq.(\ref{1/Q^4}). Since $\lambda$ increases as $\gamma_e$ decreases, we 
find that cleaner systems with smaller $\gamma_e$ get larger corrections in the CME coefficient. 
On other hand, when $\gamma_e$ approaches zero, finite systems are  in the clean limit, $l<L\ll \lambda$, so that the scale factor no longer depends on $\lambda$. In this case, all systems with different $t_z$ behave in the same way as indicated in Fig.~\ref{delta_alpha}(b) so that cleaner systems with smaller $\gamma_e$ get smaller corrections in the CME coefficient, in consistent with the previous result\cite{two-band} when $t_z=0$, there is no chiralilty imbalance and hence $\alpha = \Delta \alpha=0$.

To explore the $\beta_{{\rm CME}}$ function, $\log_{10} \Delta |\alpha|$ versus $\log_{10} L/a$ for different parameters are computed in Fig.~\ref{transition} (a) and (b). It is seen that $\beta_{{\rm CME}}$  (the slope) is always positive and approaches zero for larger systems. Since $\alpha >0$ and $\Delta \alpha<0$, it exhibits the scaling behavior of localization for the CME coefficient. Furthermore, 
Fig.~\ref{transition} (a) and (b) show that the transition from the intermediate region to the clean limit can be achieved by changing either $\gamma_e$ or $t_z$. In Fig.~\ref{transition}(b), $\gamma_e$ is fixed to $0.001$ and the transition occurs by changing $t_z$. Clearly, we see that strong suppression of the  CME coefficient occurs at $t_z=0.01$, in which  $\beta_{{\rm CME}} \sim 1$ and hence the correction, $\Delta \alpha \propto  -(L/a)$, grows.  Therefore, in the region with small $t_z$,  the Weyl semimetal is in the clean limit with
$l<L\ll \lambda$ (cf. Eq.(\ref{eta})). In this region, the CME coefficient is strongly suppressed by the quantum interference effect induced by short-range disorder.

\section{Discussion and conclusion}
In conclusion, we have investigated the effect of short-range spin-independent disorder on the chiral magnetic effects in 
Weyl semimetals. While the disorder effects on the usual conductivity in Weyl semimetals is determined by the  longitudinal part 
$\Pi_{ii}$ of the linear response,  the disorder effect on the CME coefficient is determined by the anti-symmetric combination of transverse (off-diagonal) components. The effects of disorder on conductivity include Drude-like contribution and weak localization effect and are 
analyzed based on the Kubo-Streda formula\cite{MacDonald2014,Mahan}, in which the expansion in frequency is essentially involved
due to that the electric field is related to the vector potential via $\vec{E} = - \frac{\partial \vec{A}}{ \partial t}$.
In this approach, the main contribution to the conductivity comes mainly from electrons near the Fermi surface. It is demonstrated\cite{Hikamibox} that by including the 
first weak-localization correction and two dressed Hikami boxes shown in Fig.~\ref{fig1}, the  conductivity exhibits 
weak-localization or weak-anti-localizaiton effects as observed in experiments\cite{Hikamibox, ShenPRB2015}.
On the other hand, as indicated in Eq.(12), the determination of the CME coefficient involves an additional momentum expansion due to
that the magnetic field is related to the vector potential via $\vec{B} = \nabla \times \vec{A}$. In addition,
the contribution of the disorder effect on the CME coefficient does not just come from 
electronic states near the Fermi surface. Hence the Kubo-Streda formula is not applicable. 
Furthermore, due to the anti-symmetricity of the contribution to the CME coefficient, 
two dressed Hikami boxes cancel each other out and only the first weak-localization correction contributes
the correction.   

By including the Drude corerction and computing thhe contribution due to the Cooperon channel from all electronic states via the weighting factor $W$, 
we find that while it was show that the CME effect can exist for Weyl semimetals in slowly oscillating magnetic fields,
the CME coefficient will be suppressed by short-range spin-independent disorder.
Specifically, our result shows that the quantum interference induced by short-range disorder
does not contribute the diffusive propagation, $1/(Q^2+1/\tau)$ through the Cooperon
channel in the CME coefficient. Instead, due to the additional contribution of
the wave-vector to the Cooperon propagator in the momentum expansion in
computing the CME coefficient, the Cooperon channel contributes
the CME coefficient through the quartic-momentum term, $1/(Q^2+1/\tau)^2$. As a result,
we find that the CME coefficients of finite systems manifest scaling behavior similar to scaling of 
localization even in three dimension. In particular, we find that
when the separation of Weyl nodes is small,  scaling of localization dominates and the CME coefficient is suppressed.
While the conventional view believes that the transport properties originated from topology are protected and robust against disorder, our analyses indicate that the chiral 
magnetic current can be suppressed by quantum interference due to quantum diffusion. Hence
both the chiral anomaly and the quantum interference induced by short-range disorder play 
crucial roles in the chiral magnetic effect.
Our work clarifies the role of disorder in the chiral magnetic effects and should be of help
for future investigation of elated effects in Weyl semimetals and other nodal materials.

\begin{acknowledgments}
Y. T. Lin and C. Y. Mou acknowledge support from the Ministry of Science and
Technology (MoST), Taiwan.  L. J. Zhai acknowledges support from 
National Science Foundation of China (Grant No. 11704161) and Natural Science Foundation of Jiangsu Province
of China (Grant No. BK20170309). We also acknowledge support from TCECM and Academia Sinica Research Program on
Nanoscience and Nanotechnology, Taiwan. 
\end{acknowledgments}

\appendix{}

\section{Evaluation of Cooperon Weighting Factors $W$}
\label{sec:W}

The Cooperon channel is the Cooperon propagator weighted by $W$ respectively.
In this Appendix, we derive the formulation of the weighting factors $W$ for computing
the CME correction.  According to Eq.(\ref{deltaalpha}), we need to expand $W$
as 
\begin{eqnarray}
W^{\mu'\mu}_{\nu'\nu}(\bfq)\approxeq W^{{(0)}\mu'\mu}\hspace{-1.35em}_{\nu'\nu}+qW^{^{(1)}\mu'\mu}\hspace{-1.35em}_{\nu\nu'}.
\end{eqnarray}
The formal expression of $W$ is given by Eq.({\ref{W}) with the explicit expressions for $\mathcal{G} \hat{J} \mathcal{G}$ being given by
\begin{eqnarray}
&&(\mathcal{G} \hat{J}_i \mathcal{G})_{\nu' \mu}  =  \nonumber \\
&& \mathcal{G}_{\nu'\sigma}(\bfk;i\omega_n)\hat{J}_i^{\sigma\sigma'}(\bfk)\mathcal{G}_{\sigma'\mu}(\bfk\hspace{-0.2em}+\hspace{-0.2em}\bfq ; i\omega_n\hspace{-0.2em}+\hspace{-0.2em}i\nu_m),    \nonumber\\
&&(\mathcal{G} \hat{J}_j \mathcal{G})_{\mu'\nu} = \nonumber \\ 
&& \mathcal{G}_{\mu'\lambda}(-\bfk;i\omega_n)\hat{J}_j^{\lambda\lambda'}(-\bfk)\mathcal{G}_{\lambda'\nu}(-\bfk\hspace{-0.2em}+\hspace{-0.2em}\bfq ; i\omega_n\hspace{-0.2em}+\hspace{-0.2em}i\nu_m),   \nonumber\\
\end{eqnarray}
To get the expansion of $W$, we expand the Green's functions as
\begin{eqnarray}
&&\mathcal{G}(\mathbf{k}+\mathbf{q},i\omega_n+i\nu_m) = \mathcal{G}+\mathcal{G}\bfq\cdot\nabla{h(\bfk)} \mathcal{G} +\cdots,\nonumber\\
&&\mathcal{G}(\mathbf{k}+\mathbf{q},i\omega_n) = \mathcal{G}+\mathcal{G} \bfq\cdot\nabla{h(\bfk)} \mathcal{G} \cdots,  
\end{eqnarray}
where $\nabla{h(\bfk)}$ is the gradient of Hamiltonian in Eq. (\ref{hh}). 
The summation over spin indices can be extracted out by  using Eq.(\ref{G+-}) to express the Green's function so that the spin part of $W$ can be expanded as
\begin{eqnarray} 
 (\hat{P}_w \hat{J}_i \hat{P}_v)_{\nu\mu'}(-\bfk+\bfq) &=&[\hat{N}_{i}^{wv(0)}]_{\nu\mu'}(-\bfk)  \nonumber \\
&+& q [ \hat{N}_{i}^{vw(1)} ]_{\nu\mu'} (-\bfk) + \cdots , \nonumber \\ 
(\hat{P}_s \hat{J}_j \hat{P}_t)_{\mu \nu'} (\bfk+\bfq)  &=&  [\hat{M}_{j}^{st(0)}]_{\mu\nu'}(\bfk) \nonumber \\
&+ &q[\hat{M}_{j}^{st(1)}]_{\mu\nu'}(\bfk) + \cdots.  \label{spinpart}
\end{eqnarray}
Here $\hat{P}_n$ are projection operators with $n=s,t,v$ or $w$ being band indices. $\nu,\nu',\mu$ and $\mu'$ are spin indices. 
By substituting the above expressions into Eq.({\ref{W}), the weighting factor $W$ consists of a product of four Green's functions.
To evaluate the frequency sum of the product of four Green's function, we first replace the frequency summation by a contour integral and 
 take the analytic continuation to the real frequency, $i\nu_m \to \omega+i\delta$ with  $\lim_{\omega\to 0}$. We obtain\cite{Mahan}
\begin{widetext}
\begin{eqnarray}
&&I \equiv \frac{1}{\beta}\sum_{n}\mathcal{G}^{v}(-\bfk,i\omega_n)\mathcal{G}^{w}(-\bfk,i\omega_n\hspace{-0.2em}+\hspace{-0.2em}i\nu_m)\mathcal{G}^{s}(\bfk,i\omega_n) \mathcal{G}^{t}(\bfk,i\omega_n\hspace{-0.2em}+\hspace{-0.2em}i\nu_m)\nonumber\\
&&=-2  \int^{\infty}_{-\infty} \frac{d  \varepsilon}{2 \pi} n_F(\varepsilon) \left\{ \rm{Im}  [G^v_R( -\bfk, \varepsilon) G^s_R (\bfk, \varepsilon) ] 
G^w_R(-\bfk, \varepsilon+i\delta)  G^t_R(\bfk, \varepsilon+i\delta) + \rm{Im}  [ G^w_R( -\bfk, \varepsilon) G^t_R (\bfk, \varepsilon) ] \right. \nonumber \\\
&& \hspace{10em}  \left. \times G^v_R(-\bfk, \varepsilon-i\delta)  G^s_R(\bfk, \varepsilon-i\delta), \right\} \label{spectral}
\end{eqnarray}
\end{widetext}
where $\mathcal{G}^{n}$ is the exact Green's function and $G^n_R$ is the retarded Green's function of the energy band with $n=\pm$. We shall approximate  $G^n_R$  by 
\begin{eqnarray}
G^n_R = \frac{1}{\varepsilon -E_n +\epsilon_F+ i\eta} \simeq \rm{Re} G^n_R -i \pi \delta(\varepsilon -E_n+\epsilon_F). \nonumber  \\ \label{GR}
\end{eqnarray}
Substituting Eq.(\ref{GR}) into Eq.({\ref{spectral}), we obtain
\begin{widetext}
\begin{eqnarray}
&&I =\frac{1}{2} \left\{ n_{F} (E_s-\epsilon_F)(\frac{1}{E_{s}\hspace{-0.2em}-E_{v}\hspace{-0.2em}+\hspace{-0.2em}i\eta}+\frac{1}{E_{s}\hspace{-0.2em}-E_{v}\hspace{-0.2em}-\hspace{-0.2em}i\eta})\frac{1}{E_{s}\hspace{-0.2em}-E_{w}\hspace{-0.2em}+\hspace{-0.2em}i\eta}\frac{1}{E_{s}\hspace{-0.2em}-\hspace{-0.2em}E_{t}+i\eta} \right. \nonumber \\
&&+n_{F} (E_v-\epsilon_F)(\frac{1}{E_{v}\hspace{-0.2em}-E_{s}\hspace{-0.2em}+\hspace{-0.2em}i\eta}+\frac{1}{E_{v}\hspace{-0.2em}-E_{s}-i\eta})\frac{1}{E_{v}\hspace{-0.2em}-E_{w}\hspace{-0.2em}+\hspace{-0.2em}i\eta}\frac{1}{E_{v}\hspace{-0.2em}-\hspace{-0.2em}E_{t}+i\eta}  \nonumber \\
&&+n_{F}{(E_t-\epsilon_F\hspace{-0.2em})}  (\frac{1}{E_{t}\hspace{-0.2em}-E_{w}\hspace{-0.2em}+\hspace{-0.2em}i\eta}+\frac{1}{E_{t}\hspace{-0.2em}-E_{w}-i\eta})\frac{1}{E_{t}\hspace{-0.2em}-E_{v}\hspace{-0.2em}+\hspace{-0.2em}i\eta}\frac{1}{E_{t}\hspace{-0.2em}-\hspace{-0.2em}E_{s}+i\eta} \nonumber \\
&& \left. +n_{F}{(E_w-\epsilon_F)}(\frac{1}{E_{w}\hspace{-0.2em}-E_{t}\hspace{-0.2em}+\hspace{-0.2em}i\eta}+\frac{1}{E_{w}\hspace{-0.2em}-E_{t}-i\eta})\frac{1}{E_{w}-\hspace{-0.2em}E_{v}+\hspace{-0.2em}i\eta}\frac{1}{E_{w}-\hspace{-0.2em}E_{s}+\hspace{-0.2em}i\eta} \right\},\nonumber \\ \label{GGGG}
\end{eqnarray}
\end{widetext}
where $n_F$ is the Fermi-Dirac distribution function. By combing Eqs.(\ref{spinpart}) and (\ref{GGGG}) with Eq.(\ref{W}), the weight factor 
$W^{{(0)}\mu\mu'}\hspace{-1.35em}_{\nu\nu'}$  can be expressed into the following forms
\begin{eqnarray}
&&\hspace{-3em}\left[ W^{{(0)}}_{xy}\right]^{\mu\mu'}_{\nu\nu'}\hspace{-0.5em}=\hspace{-0.3em}\frac{e^2}{\hbar^2}
\begin{pmatrix}    
R_{1} & 0 & 0 & 0 \\
0 & R_{2} & R_{3} & 0 \\
0 & R_{4} & R_{5} & 0 \\
0 & 0 & 0 & R_{6}
\end{pmatrix}, \nonumber \\
&&\hspace{-3em}\left[ W^{{(0)}}_{yz}\right]^{\mu\mu'}_{\nu\nu'}\hspace{-0.5em}=\hspace{-0.3em}\frac{e^2}{\hbar^2}
\begin{pmatrix}    
0        & R_{7} &  R_{8}    & 0 \\
R_{9} & 0        &  0           & R_{10} \\
R_{11} & 0       &  0           & R_{12} \\
0       & R_{13} &  R_{14} & 0
\end{pmatrix}, 
\nonumber \\
&&\hspace{-3em}\left[ W^{{(0)}}_{zx}\right]^{\mu\mu'}_{\nu\nu'}\hspace{-0.5em}=\hspace{-0.3em}\frac{e^2}{\hbar^2}
\begin{pmatrix}    
0 & R_{15} & R_{16} & 0 \\
R_{17} & 0 & 0 & R_{18} \\
R_{19} & 0 & 0 & R_{20} \\
0 & R_{21} & R_{22} & 0
\end{pmatrix},
\nonumber \\ \label{Wcomp}
\end{eqnarray}
where $R_n$ with $n=1,2,3,\cdots,22$ are non-vanishing elements determined by $\hat{N}_{i}^{0)}$ and $\hat{M}_{i}^{(0)}$.

Similarly, the linear $q$ term of the weighting factor is given by
\begin{eqnarray}
&&\hspace{-2em}W^{(1)\mu\mu'}\hspace{-1.3em}_{\nu\nu'} = \nonumber\\
&&\frac{1}{3\beta\Omega}\sum_{n,\bfk}   \left[(\mathcal{G} \hat{J}_y \mathcal{G})_{\nu \mu'}(\mathcal{G}\nabla\cdot{h(\bfk)}\mathcal{G} \hat{J}_x \mathcal{G})_{\mu\nu'}(\bfk;i\omega_n,i\nu_m)\right. \nonumber\\
&&\hspace{2em}\left. +(\mathcal{G} \hat{J}_y \mathcal{G}\nabla\cdot{h(\bfk)}\mathcal{G})_{\nu \mu'}(\mathcal{G} \hat{J}_x \mathcal{G})_{\mu\nu'}(\bfk;i\omega_n,i\nu_m)  \right]\nonumber\\
&&\hspace{2em}-(x \leftrightarrow y)   \nonumber\\
&&\hspace{-3em}=\frac{1}{3\Omega}\sum_{\bfk}\hspace{-0.5em}\sum_{stvwr=\pm}\left[[\hat{N}_{x}^{vw(0)}]_{\nu\mu'}\hspace{-0.2em}\otimes[\hat{M}_{y}^{ts(1)}]_{\mu\nu'}-(x \leftrightarrow y)\right]\nonumber\\
&&\hspace{2em}\times\frac{1}{\beta}\sum_{n}(\mathcal{G}^{v} \mathcal{G}^{w} )(\mathcal{G}^{r} \mathcal{G}^{s} \mathcal{G}^{t} )(\bfk;i\omega_n,i\nu_m)\nonumber\\
&&\hspace{-2em}+\frac{1}{3\Omega}\sum_{\bfk}\hspace{-0.5em}\sum_{stvwr=\pm}\left[[\hat{N}_{x}^{vw(1)}]_{\nu\mu'}\hspace{-0.2em}\otimes[\hat{M}_{y}^{ts(0)}]_{\mu\nu'}-(x \leftrightarrow y)\right]\nonumber\\
&&\hspace{2em}\times\frac{1}{\beta}\sum_{n}(\mathcal{G}^{v} \mathcal{G}^{w} \mathcal{G}^{r} )(\mathcal{G}^{s} \mathcal{G}^{t} )(\bfk;i\omega_n,i\nu_m).\nonumber\\
\end{eqnarray}
Following the same procedure,  $W^{(1)\mu\mu'}\hspace{-1.3em}_{\nu\nu'}$ can be obtained.  However, according to Eq.(\ref{deltaalpha}), $W^{(1)}$ couples to $\Gamma^{(0)}$. Hence $W^{(1)}$ contributes a sub-leading term, $1/Q^2$, and can be neglected.


\begin{thebibliography}{99}
\bibitem{Geim1} K.S. Novoselov et al., Science {\bf 306}, 666 (2004).
\bibitem{Geim2}  A.H. Castro Neto, F. A. H. Guinea, N. M. R. Peres,  K. S. Novoselov, and
A. K. Geim, Rev. Mod. Phys. {\bf 81}, 109 (2009).
\bibitem{Kane} M. A. Hasan and C. L. Kane, Rev. Mod. Phys. {\bf 82}, 3045 (2010).
\bibitem{Zhang} X. L. Qi and S. -C.  Zhang, Rev. Mod. Phys. {\bf 83}, 1057 (2011).
\bibitem{Murakami} S. Murakami, New J. Phys. {\bf 9}, 356 (2007).
\bibitem{Chou} Po-Hao Chou, Liang-Jun Zhai, Chung-Hou Chung, Chung-Yu Mou, and Ting-Kuo Lee, Phys. Rev. Lett. 116, 177002 (2016).
\bibitem{Na3Bi} Z. K. Liu et. al., Science {\bf 343}, 864 (2014).
\bibitem{Cd2As3_1} S. Borisenko, Q. Gibson, D. Evtushinsky, V. Zabolotnyy, B. Buchner, and R.J. Cava, Phys. Rev. Lett. {\bf 113}, 027603 (2014).
\bibitem{Cd2As3_2} M. Neupane, Nature Communications {\bf 5}, 3786 (2014).
\bibitem{TaAs1} S.-Y. Xu et al., Science {\bf 349}, 613 (2015).
\bibitem{TaAs2} B.Q.Lv, H.M.Weng, B.B.Fu, X.P.Wang, H.Miao, J.Ma, P. Richard, X.C.Huang, L.X.Zhao, G.F.Chen, Z.Fang, X.Dai, T. Qian, and H.Ding ,Phys. Rev. X {\bf 5}, 031013 (2015)
\bibitem{review} For recent reviews, see P. Hosur and X. L. Qi, C. R. Physique {\bf 14}, 857 (2013) ;D. E. Kharzeev, Progr. Part. Nucl. Phys. {\bf 75}, 133 (2014); A. A. Burkov, J. Phys. Condens. Matter {\bf 27}, 113201 (2015).
\bibitem{Zhai}Liang-Jun Zhai, Po-Hao Chou, and Chung-Yu Mou, Phys. Rev. B 94, 125135 (2016).
\bibitem{Nielsen}
H.B. Nielsen and M. Ninomiya, Nucl. Phys. B 185, 20 (1981).
\bibitem{Burkov2015}
A.A. Burkov, J. Phys.: Condens. Matter 27, 113201, (2015).
\bibitem{Burkov2012}
A.A. Zyuzin, A.A. Burkov, Phys. Rev. B 86, 115133  (2012)
\bibitem{two-band} M.C. Chang, M.F. Yang, Phys. Rev. B {\bf 91}, 115203 (2015). Note that an extra minus in front of $\alpha$ in this paper should be absent.
\bibitem{Basar} G. Basar, D. E. Kharzeev, and Ho-Ung Yee, Phys. Rev. B {\bf 89}, 035142 (2014).
\bibitem{Franz} M.M. Vazifeh and M. Franz, Phys. Rev. Lett. {\bf 111}, 027201 (2013).
\bibitem{Yamamoto} N. Yamamoto, Phys. Rev. D {\bf 92}, 085011 (2015).
\bibitem{Scattering theory} P. Baireuther, J.A. Hutasoit, J.Tworzydlo and C.W.J Beenakker, New J. Phys. {\bf 18}, 045009 (2016).
\bibitem{gyro} S. Zhong, J. E. Moore, and I. Souza, Phys. Rev. Lett. {\bf 116}, 077201 (2016).
\bibitem{gyro1} P. Goswami, G. Sharma, and S. Tewari, Phys. Rev. B {\bf 92}, 161110(R), (2015).
\bibitem{gyro2} A. Sekine and K. Nomura, Phys. Rev. Lett. {\bf 116}, 096401, (2016).
\bibitem{gyro3} J. Ma and D. A. Pesin, Phys. Rev. Lett. {\bf 118}, 107401, (2017).
\bibitem{neg}Xiaochun Huang, Lingxiao Zhao, Yujia Long, Peipei Wang, Dong Chen, Zhanhai Yang, Hui Liang, Mianqi Xue, Hongming Weng, Zhong Fang, Xi Dai, and Genfu Chen, Phys. Rev. X {\bf 5}, 031023  (2015).
\bibitem{Hikamibox} E. McCann et al., Phys. Rev. Lett. {\bf 97}, 146805 (2006); H.-Z. Lu, J. Shi, and S.-Q. Shen, Phys. Rev. Lett. {\bf 107}, 076801
(2011).
\bibitem{ShenPRB2015}
H.Z. Lu and S.Q. Shen,  Phys. Rev. B {\bf 92}, 035203  (2015).
\bibitem{MacDonald2014}
Y. Araki, G. Khalsa, A.H. MacDonald Phys. Rev. B 90, 125309 (2014)
\bibitem{tewari} P. Goswami and S. Tewari, arXiv:1311.1506.
\bibitem{Akkermans}
E. Akkermans and G. Montambaux, {\it Mesoscopic Physics of Electrons and Photons }, Cambridge University Press (2007).
\bibitem{Mahan}
Gerald D. Mahan, {\it Many-Particle Physics }, Springer; p. 619, 3nd ed. (1990).














\end{thebibliography}
 \end{document}